\documentclass[useAMS,usenatbib]{mn2e}

\usepackage{graphicx}
\usepackage{txfonts}
\usepackage[authoryear]{natbib}
\usepackage{subfigure}

\title[Non-Keplerian RV effects in binary stars]{Non-Keplerian effects in precision radial velocity measurements of double-line spectroscopic binary stars: numerical simulations}
\author[P. Sybilski et al.]
{P. Sybilski,$^{1}$\thanks{E-mail:sybilski@ncac.torun.pl}
M. Konacki,$^{1,2}$ S. K. Koz{\l}owski,$^{1}$ K. G. He{\l}miniak$^{3,1}$\\
$^{1}$Nicolaus Copernicus Astronomical Center, Polish Academy of Sciences, Rabia\'nska 8, 
     87-100 Toru\'n, Poland\\
$^{2}$Astronomical Observatory, Adam Mickiewicz University, S{\l}oneczna 36, 60-186 Pozna\'n, Poland\\
$^{3}$Departamento de Astronom\'{i}a y Astrof\'{i}sica, Pontificia Universidad Cat\'{o}lica, Vicu\~{n}a Mackenna 4860, 782-0436 Macul, Santiago, Chile\\}

\begin{document}

\date{Accepted ... Received ...; in original form ...}

\pagerange{\pageref{firstpage}--\pageref{lastpage}} \pubyear{2012}

\maketitle

\label{firstpage}

\begin{abstract}
Current precision in radial velocity (RV) measurements of binary stars reaches $\sim$2 ms$^{-1}$. 
This level of precision means that RV models have to take into account additional non-Keplerian effects such as tidal and rotational distortion of the components of a binary star, relativistic effects and orbital precession. Such an approach is necessary when one wants to search for planets or precisely measure fundamental parameters of stars with a very high accuracy using precision RVs of binary stars.

We generate synthetic binaries using Yonsei-Yale stellar models. For typical representatives we investigate the impact of various orbital orientations and different non-Keplerian effects on the RV curves. To this end we simulate RV observations with an added white noise of different scale. Subsequently we try to reconstruct the input orbital parameters and their errors by fitting a model using a standard least-squares method. In particular we investigate the connection between the tidal distortion of the
shape of the stars and the best-fit orbital eccentricity, the possibility of deriving orbital inclination of a non-eclipsing binary star by exploiting relativistic effects and the circumstances in which the orbital precession can be detected.

We confirm that the method proposed by \cite{zucker07} to obtain orbital inclination with use of the relativistic effect does work in favourable cases and that it can be used even for orbital configurations far from an edge-on orientation. We show that the RV variations imposed by tidally 
distorted stars can mimic non-zero eccentricity in some binaries. The scale of such an effect depends on the RV accuracy. Finally, we demonstrate that the apsidal precession can be easily detected with precision RVs. In particular we can detect orbital precession of $10^{-4}$ rad yr$^{-1}$, $10^{-3}$ rad yr$^{-1}$ for precision of RVs of 1 ms$^{-1}$ and 10 ms$^{-1}$ respectively.
\end{abstract}


\section{Introduction}
\label{sec1}
In 1842, Christian Andreas Doppler published his work ``On the coloured light of the binary stars and some other stars of the heavens" \citep{doppler1842}. Although he did not get it quite right then, his work has started the career of the well known Doppler effect. Over a century has passed and the best radial velocity (RV) precision for the primaries of double-lined spectroscopic binaries remained almost unchanged at about 100 ms$^{-1}$ until recently. In 2003, Zucker et al. presented a precise radial velocity measurement of a binary component with precision reaching 10 ms$^{-1}$. With the use of TODCOR, high quality echelle spectra from CORALIE and faint second component not interfering with the lines of primary component the new capability was reached. A year later Skuljan et al (2004) have reached a precision of 14 ms$^{-1}$ in a similar configuration. Another important step was development of spectra disentangling code by M. Konacki utilizing the iodine cell technique and making the regular, high precision RV measurement of many binaries, with two similar components, possible. It is now within the reach of many instruments to achieve the precision of up to a few meters per second for the components of double-lined spectroscopic binary stars \citep{konacki10}.

The main goal of this paper is to investigate the selected features of double-lined binary stars, their potential and problems in the context of precise RV measurements. We based our selection of effects on the recent publications by \cite{konacki10} and \cite{mazeh08} and observational practices. In the new precision regime the tidal effects become important. Stars cannot be treated as single points in space. The presence of another body in an orbit around a star generates tidal effects influencing the shape of a star. Because of it, the measured radial velocity is different from that derived from simple Keplerian motion \citep{wilson76, kopal80}. This effect, when neglected, may be responsible for the measurement of a non-zero eccentricity in an otherwise circular orbit. Given sufficient precision of the RV curve, one can rule out this situation to a certain extent. Moreover the relativistic effects may be a source of additional information about a system. Therefore, we explore the relations derived by \cite{zucker07} relating the relativistic effects and the orbital inclination, $i$. With sufficient precision of the RV measurements, one can derive $i$ of the system even for a non-eclipsing systems and using only RVs. The last effect we examine is the apsidal precession. We determine the required RV precision to measure the rate of apsidal precession. To achieve these goals we conduct thorough numerical simulations using currently available codes for modelling binary stars. Our analysis is aimed to support precise, current and future, RV surveys of binary stars by investigating the challenges and possibilities.

In section 2, we describe models as well as a synthetic population of binary stars which are used in the tests. This section also explains an algorithm employed to obtain a final set of parameters based on artificial data. In Section 3, we analyse the impact of tidal effects on the RV curve and eccentricity. In Section 4, we test the idea proposed by \cite{zucker07} of using pure RV data as an independent source of the orbital inclination. We test its feasibility and usability. In Section 5 we check the limits of apsidal motion detection for different RV precision. The conclusions are provided in Section 6.

\section{Models}
\label{sec2}
The tidal and rotational effects were studied and implemented as a computer algorithm by \cite{wilson93}. The analytical approximation was provided by \cite{kopal80}. We decided to use WD code based on the models of \cite{wilson71} and \cite{wilson79} as the source of synthetic RV data. The WD code includes all the effects we are interested in and much more. It is moreover a widely accepted tool for modelling eclipsing binary stars. 

We prepared the artificial population of binary stars in different orbital configurations, containing from 10000 to 400000 objects. The resolution of such a grid and the choice of parameters depended on an analysed problem and a computing time necessary to pass the grid. Every problem had been computed several times to confirm the results. The input parameters are based on the widely used Yonsei-Yale ($Y^2$) isochrones \citep{demar04, yi01}. We tested other ones like PADOVA \citep{girardi00, marigo08} or GENEVA \citep{lejune01} but the resulting differences had minor impact on the results and none on the conclusions. Chemical abundances in our set are described by X = 0.71, Y = 0.27, Z = 0.02, this composition is similar to the Sun. Certain values for the test stars were obtained by a linear interpolation of neighbouring values from the $Y^2$ table. Additional information necessary to describe the components of a binary, like the gravitational brightening coefficients, were taken from  \cite{alencar97, alencar99, claret00a, claret00b}. Stars and their orbital parameters are described in Table \ref{tab:01_StarParameters} and \ref{tab:02_OrbitParameters} respectively. Limb darkening was computed with cosine law with WD code \citep{wilson07}. Stellar atmosphere formulation was used for local emission on binary star components, instead of a black body approximation. Other internal parameters of WD code used in our simulations are summarised in Table \ref{tab:03_WdParameters}. We assume constant $F$, periastron-synchronised ratio of the axial rotation rate to the mean orbital rate. The sample aims to explore the set of close binaries with masses between 0.5 M$_\odot$ and 1.5 M$_\odot$. In this range the examined effects are important for precise RV measurements. Symbols used in the tables are the orbital elements, $e$ for the eccentricity, $\omega$ for the argument of pericenter. The parameter $\sigma_{\mathrm{RV}}$ describes the white noise added to every measurement and $i$ is given in degrees. In the case of an eclipsing configuration, we chose orbits close to an edge-on orientation, $i$ ranges from 87$\degr$ to 93$\degr$. This range assures that every object has an eclipse. For a given range of orbital inclinations (eclipsing or not), the configurations were drawn from a uniform distribution of the orbital inclination. \textbf{This distribution differs from the isotropic distribution of possible orbits in space. The results are not a representative statistical sample of real binaries, but rather demonstrate the kind of effects we expect to see in the sampled range of parameters.}

In our simulations we proceeded as follows. We generated a synthetic population of binaries and for each
binary we computed a synthetic RV set with white noise added to mimic a measurement error. For every binary,
we computed thirty RVs for the primary and secondary components randomly spread over a given time span of a data set. Such an RV set was used to fit a model. We used the MINPACK library \citep[][]{more80, more84a}\footnote{www.netlib.org} to carry out a least-squares fitting. As input parameters we used the original ones
modified by at least half of a percent.

In the models we assume that the stars are separated and none of the components fills its Roche lobe. This is accomplished by testing the WD code output and fulfilling the inequality:
\begin{equation}
\label{eq01:DetachedCriterion}
2 (R_1 + R_2) < a (1 - e)
\end{equation}
where $R_1$, $R_2$ are the radii of the first and second star respectively. This assures that we deal only with detached systems. We analyse RV curves for both stars at the same time. Moments of eclipses are avoided. It is a common approach made by observers which makes the RV modelling much easier. However, it is worth noticing that eclipses may be a reliable source of additional information especially in the case of rapidly rotating non-circular binaries where the non-Keplerian effects have a high amplitude. This is known as the Rossiter-McLaughlin effect \citep[][]{rossiter24, mcLaughlin24}. Unfortunately WD code does not allow to model a misalignment between the star's spin and the orbital angular momentum of the binary, and covers only the basic case of aligned rotation. In our simulations we used as an input only the aligned cases. Therefore our results are not influenced by this limitation of the WD code. However during the normal analysis of observational data it should be included. Proper modelling of spectral lines and radial velocity should take this effect into account, especially in the case of eclipses, where it is essential. Omitting it may cause the observed precession rate to be four times slower than the theoretical rate \citep{albrecht09}. The misalignment will change the shape of the RV curve but it's highest amplitude will remain the same. Misaligned configurations produce generally lower amplitudes.

Analytical approach to the RV curves was described by \cite{kopal80, kopal80b}. More recent publications regarding the subject by \cite{zucker07} and \cite{mazeh08} prove the significance of modelling close interactions in binaries. Development of high resolution spectrographs like HARPS \citep{rupprecht04}, HIRES \citep{vogt94} or the incoming ESPRESSO \citep{pepe10} and methods like iodine cell calibration \citep{konacki05} ensure the advancement of precision in RV measurements. Software and model evolution is also needed to fully utilise hardware improvements. One example of such an improvement is the spectral disentangling method \citep[][]{hadrava95, konacki10}. 

\begin{table*}
\caption{Stellar parameters from $Y^2$ models for the components of synthetic binaries. Permutations used in our simulation assume the same age of components.}
\label{tab:01_StarParameters}
\centering
\begin{tabular}{r  c  c  c  c  c  c  c  c  l}
\hline
Parameter & Star 1 & Star 2 & Star 3 & Star 4 & Star 5 & Star 6 & Star 7 & Star 8 & Unit \\
\hline \hline
Age  & 0.8 & 0.8 & 0.8 & 3 & 3 & 3 & 12  & 12  &    Gyr \\
Mass & 0.5 & 1 & 1.5 & 0.5 & 1 & 1.5 & 0.5 & 1 & M$_\odot$ \\
Log g & 4.78 & 4.53 & 4.22 & 4.77 & 4.49 & 3.34 & 4.75 & 3.68 & - \\
Temperature & 3521 & 5619 & 6995 & 3518 & 4927 & 5676 & 3563 & 4823 & K \\
Gravitational brightening & 0.12 & 0.396 & 0.272 & 0.12 & 0.396 & 0.4 & 0.12 & 0.38 & - \\
\hline
\end{tabular}
\end{table*}

\begin{table}
\caption{Orbital and other parameters used in the simulations.}
\label{tab:02_OrbitParameters}
\centering
\begin{tabular}{r c l}
\hline
Varied Parameter & Values Range & Unit \\
\hline
\hline
$e$ & 0.0, 0.1, .. , 0.9 & - \\
$\omega$ & 0.0 - 2$\pi$ & - \\
$\sigma_{\mathrm{RV}}$ & 1, 10, 100 &ms$^{-1}$ \\
$P$ & 3, 5, 10 & day \\
$i$ (eclipsing binaries) & 87, 88 .. 93 & $\degr$ \\
$i$ (relativistic effects) & 0, 3 .. 90 & $\degr$ \\
$i$ (apsidal precession) & 87 & $\degr$ \\
\hline
\end{tabular}
\end{table}

\begin{table}
\caption{Internal parameters of Wilson-Devinney code used in the simulations.}
\label{tab:03_WdParameters}
\centering
\begin{tabular}{r c}
\hline
Parameter & Value \\
\hline
\hline
Grid size for star 1 and 2 & 30 \\
Bolometric albedo of star 1 and 2 & 0.5 \\
$F$ & $\sqrt{\frac{1+e}{(1-e)^3}}$ \\
\hline
\end{tabular}
\end{table}

\section{Tidal effect and eccentricity}
\label{sec3}
\begin{figure*}[ht]
\centering
\subfigure[]{
\includegraphics[width=\columnwidth]{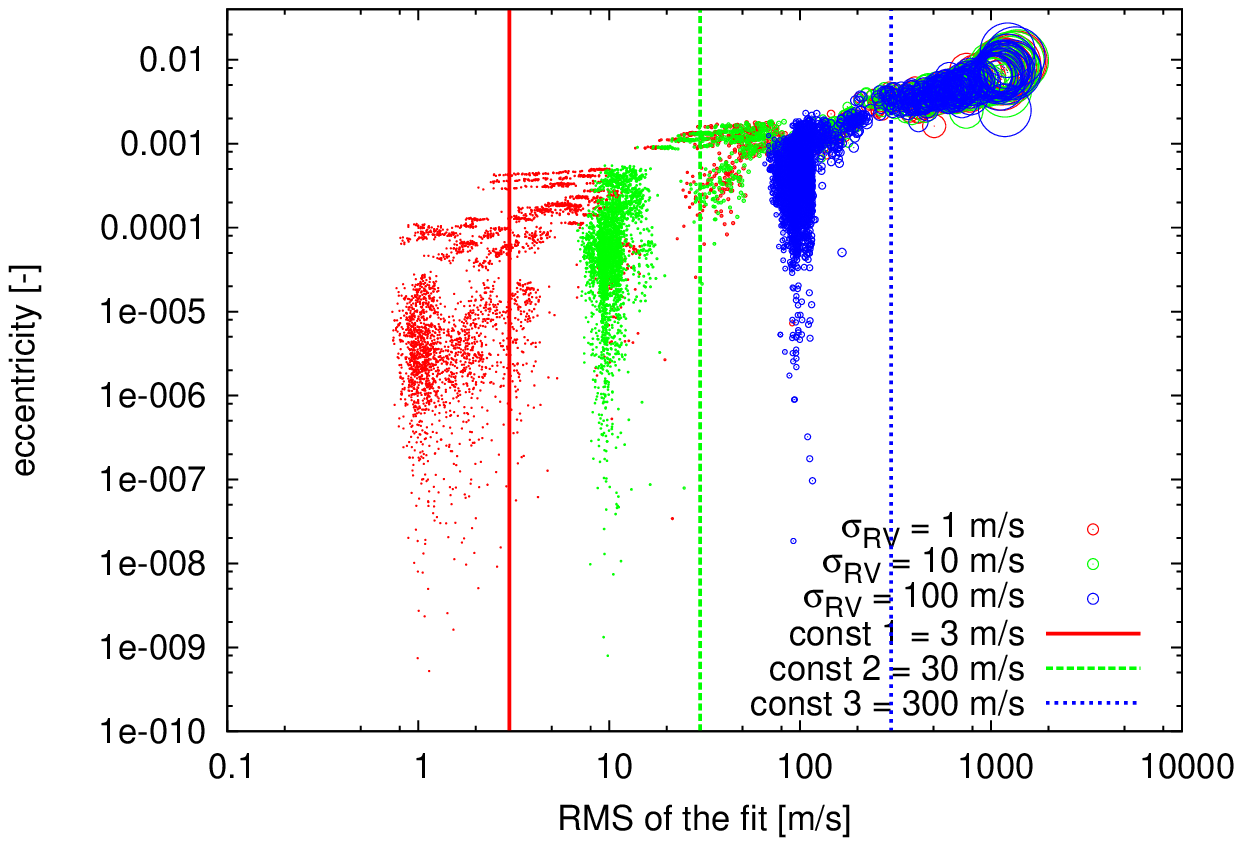}
\label{a1}
}
\subfigure[]{
\includegraphics[width=\columnwidth]{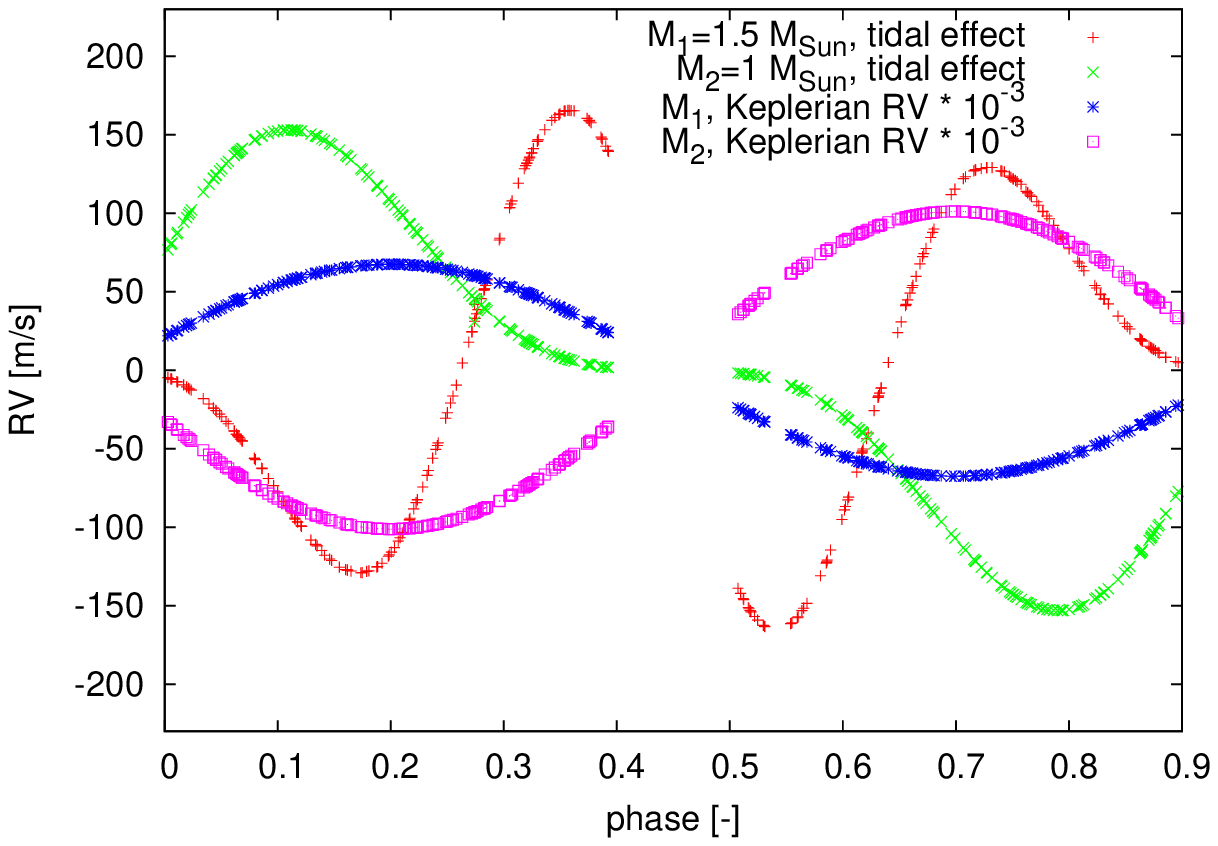}
\label{a2}
}
\subfigure[]{
\includegraphics[width=\columnwidth]{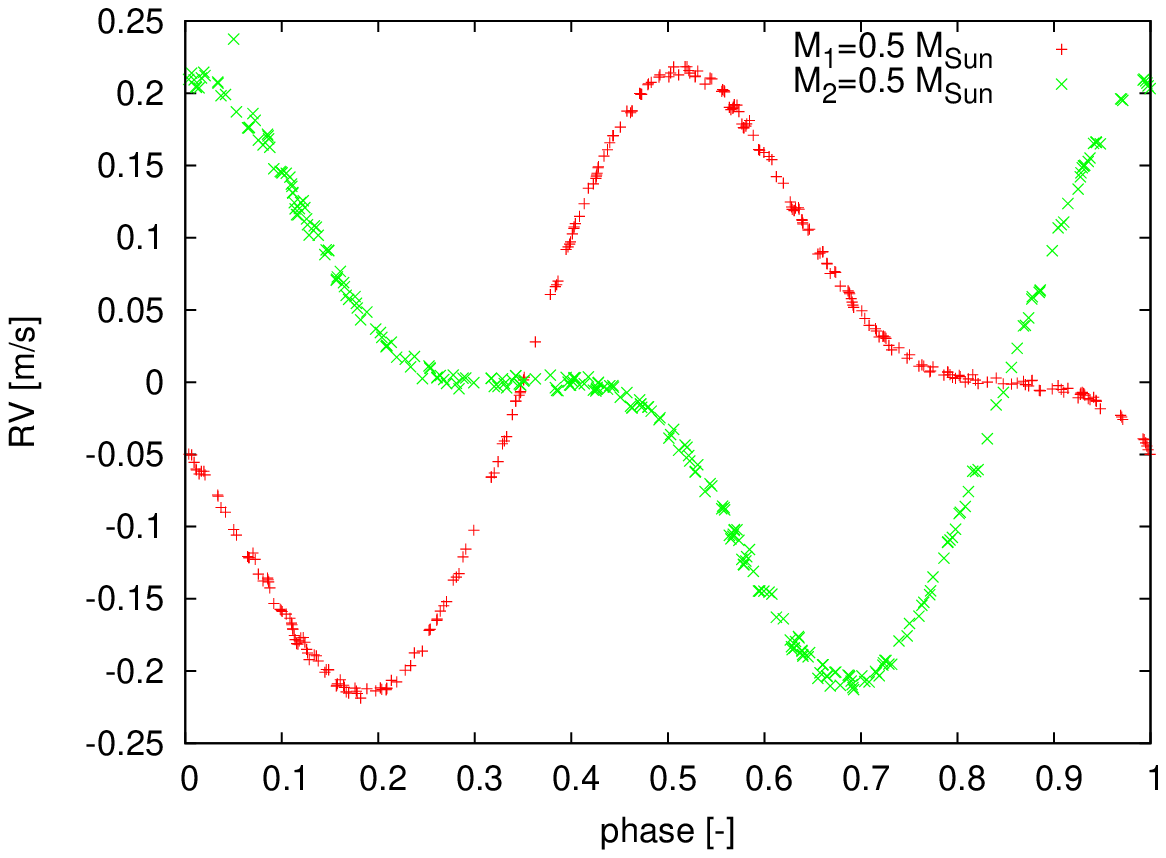}
\label{a3}
}
\subfigure[]{
\includegraphics[width=\columnwidth]{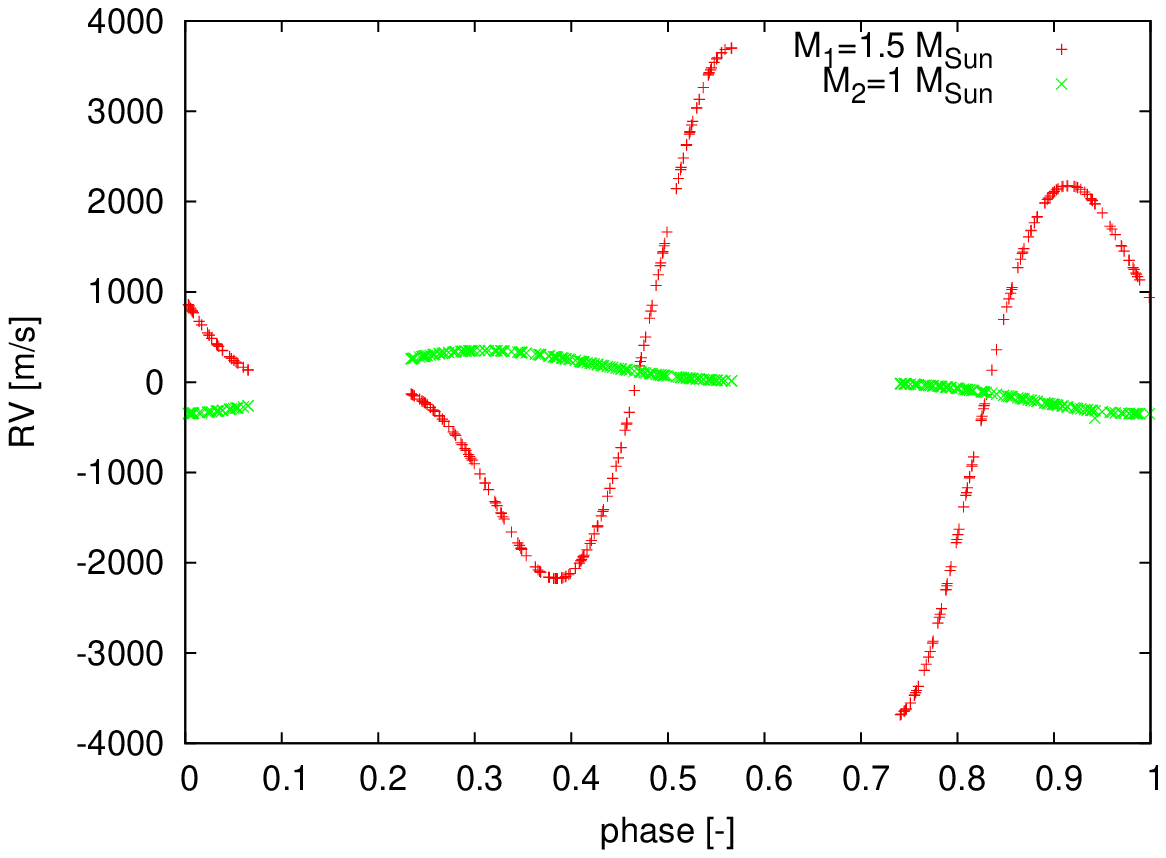}
\label{a4}
}
\caption{Illustration of the tidal effect's impact on fitting just a simple Keplerian model to the RVs of a binary star. The panel (a) shows the full result of our simulation, a relation between the best-fit RMS of the final fit and the eccentricity. Synthetic data sets are based on WD code while a simple Keplerian model is used for modelling. Vertical lines denote three constant RMS intervals 3, 30 and 300 ms$^{-1}$. All the $e$ results on the left side of each line may be easily taken as a genuine eccentricity. The size of each dot is determined by the RMS of the fit. The panels (b-d) show three typical examples of tidal effect in RV: average, minimal and maximal. As one may easily predict, more massive, tight and old binaries tend to have much bigger surface distortions, and because of this, higher semi-amplitudes of the tidal effect. Gaps in the curves are caused by eclipses where no measurements were modelled. The system shown in the panel (c) has two shallow and short eclipses around phase 0.3 and 0.95, due to the orbital configuration and small size of the components. Panel (b) shows additional information, scaled down by factor of 1000 RV coming from Keplerian motion of components around common center of mass.}
\label{fig:examplesTides}
\end{figure*}

The shape of stars in close binaries differs from standalone objects. Both components are subject to mutual tidal forces and also mutual irradiation. Deviation from the spherical shape and non-uniform surface brightness are a source of an RV curve distortion. An average amplitude of the tidal effect in our sample is 174 ms$^{-1}$, but can be as big as 3800 ms$^{-1}$ or as small as 0.25 ms$^{-1}$. We define tidal effect semi-amplitude as the maximal difference between the Keplerian RV and the RV curve of the same star with the tidally distorted surface. In our simulations this parameter is estimated on the base of randomly distributed measurements across the system's period. Examples of three cases are presented in Figure \ref{fig:examplesTides}, panels (b-d) respectively. When this effect is ignored one can obtain a best-fit RV root mean square (RMS) of 1 ms$^{-1}$ or better, only in the case of well separated objects. 

Moreover, the tidal effect may mimic an eccentric orbit when a simple Keplerian model is fitted to the RV data. The tidal effect is often easily incorporated into the eccentricity. We show to what extent the tidal effect may produce correct fits with  RMS of the best-fit $\sigma_{fit}$ lower than $3\sigma_{\mathrm{RV}}$. To this end, almost 12000 systems had been tested. The synthetic population consists of all age and mass permutations from Table \ref{tab:01_StarParameters}, used with the entire range of varied orbital parameters from Table \ref{tab:02_OrbitParameters}. Only the eccentricity $e$ is kept at zero. 

\begin{figure}
\includegraphics[scale=0.65]{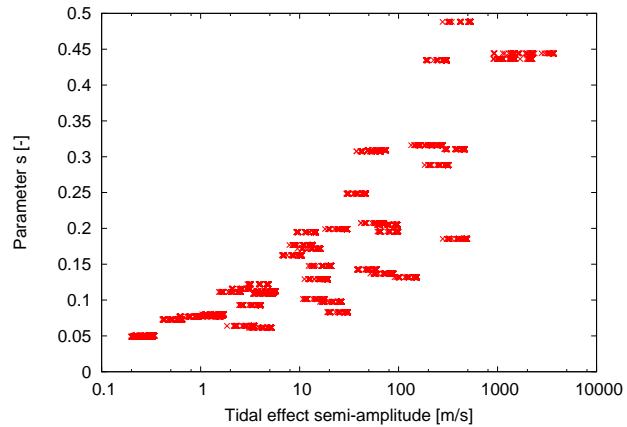}
\caption{Illustration of the relation between the semi-amplitude of the tidal effect and the $s$ parameter. Almost 12000 simulations in this figure let us estimate the safe limits for avoiding the tidal effect at a given RV precision.}
\label{fig:sVsAmp}
\end{figure}

From our simulations emerges a safe limit for a separation of stars: $a(1-e)>20(R_1+R_2)$. In this regime, we may safely ignore the tidal effects and assume them to contribute less than 0.1 ms$^{-1}$. The tides' impact on the RVs starts to drop rapidly with the parameter $s$ lower than 0.1, where $s$ is given by the equation:
\begin{equation}
\label{eq04:sParameter}
s = \frac{R_1+R_2}{a(1-e)}
\end{equation}
A detailed relation between the $s$ parameter and the semi-amplitude of the tidal effect is illustrated in Figure \ref{fig:sVsAmp}. 

\begin{figure*}
\centering
\subfigure[]{
\includegraphics[width=\columnwidth]{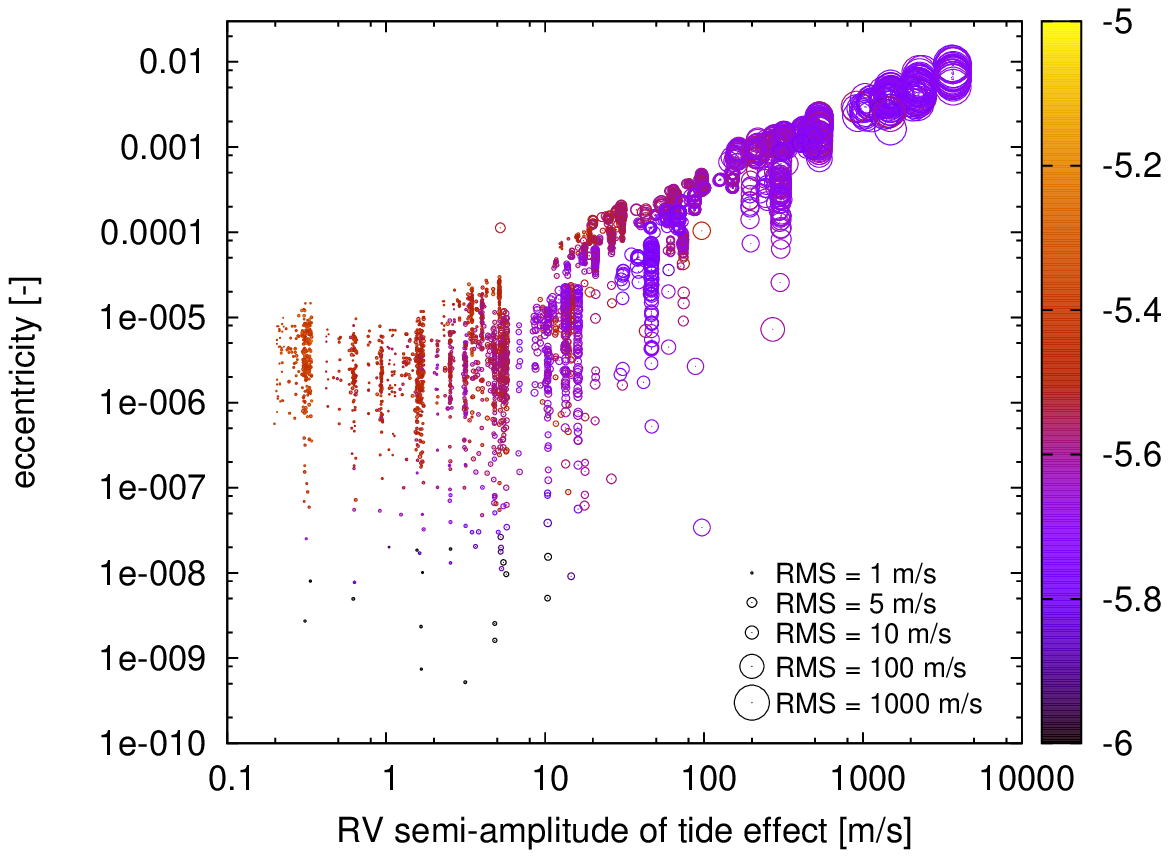}
}
\subfigure[]{
\includegraphics[width=\columnwidth]{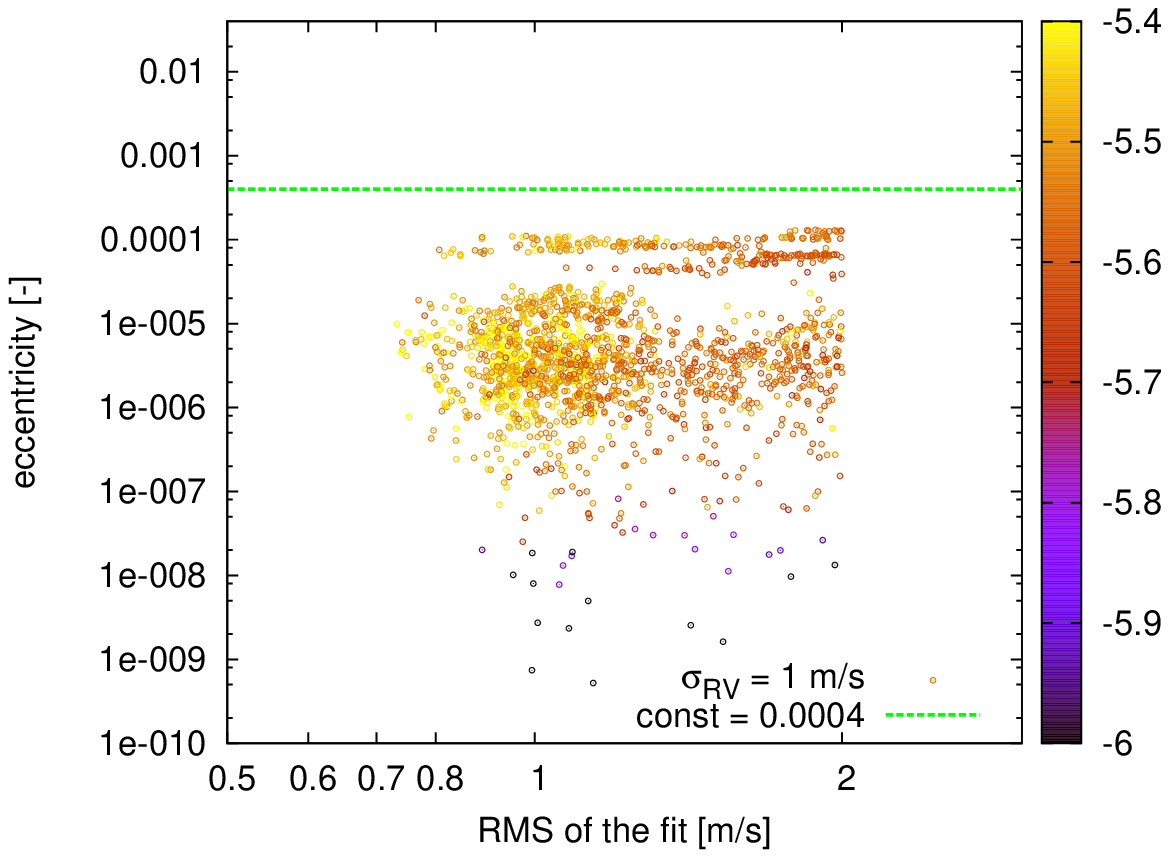}
}
\subfigure[]{
\includegraphics[width=\columnwidth]{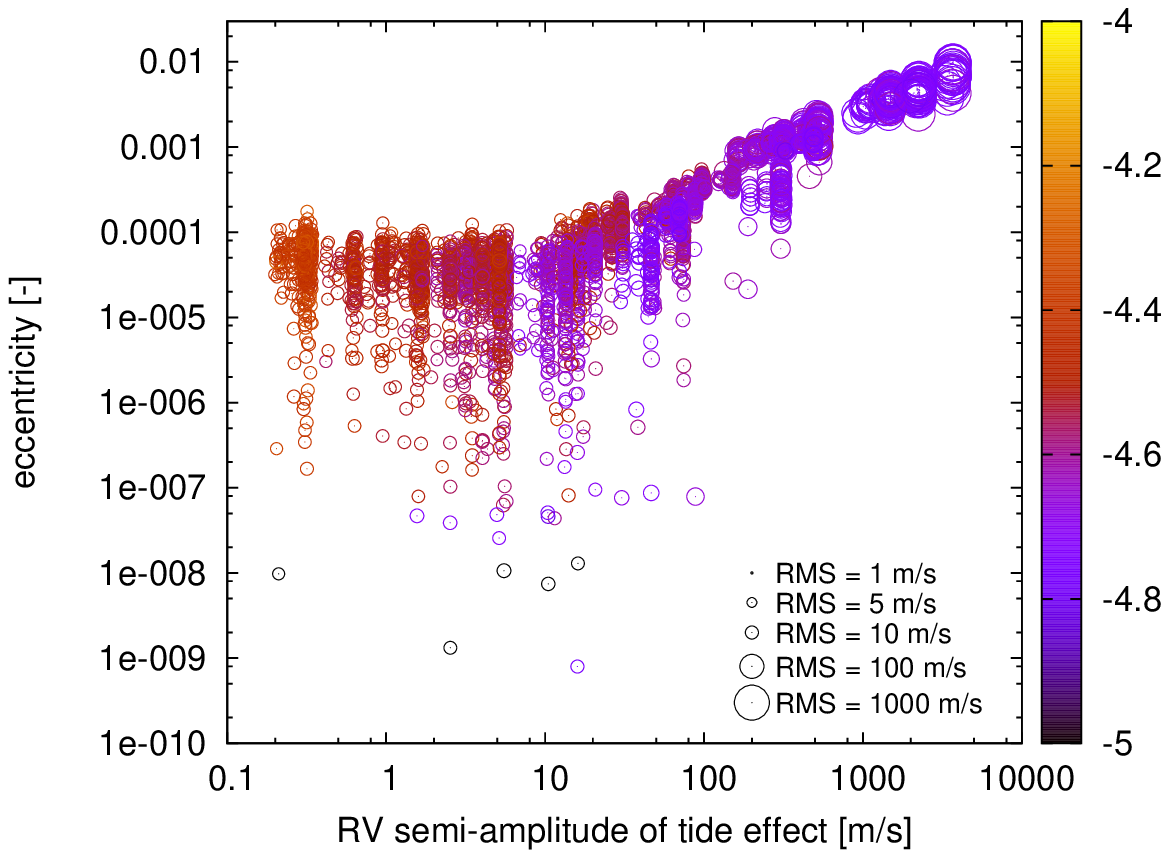}
}
\subfigure[]{
\includegraphics[width=\columnwidth]{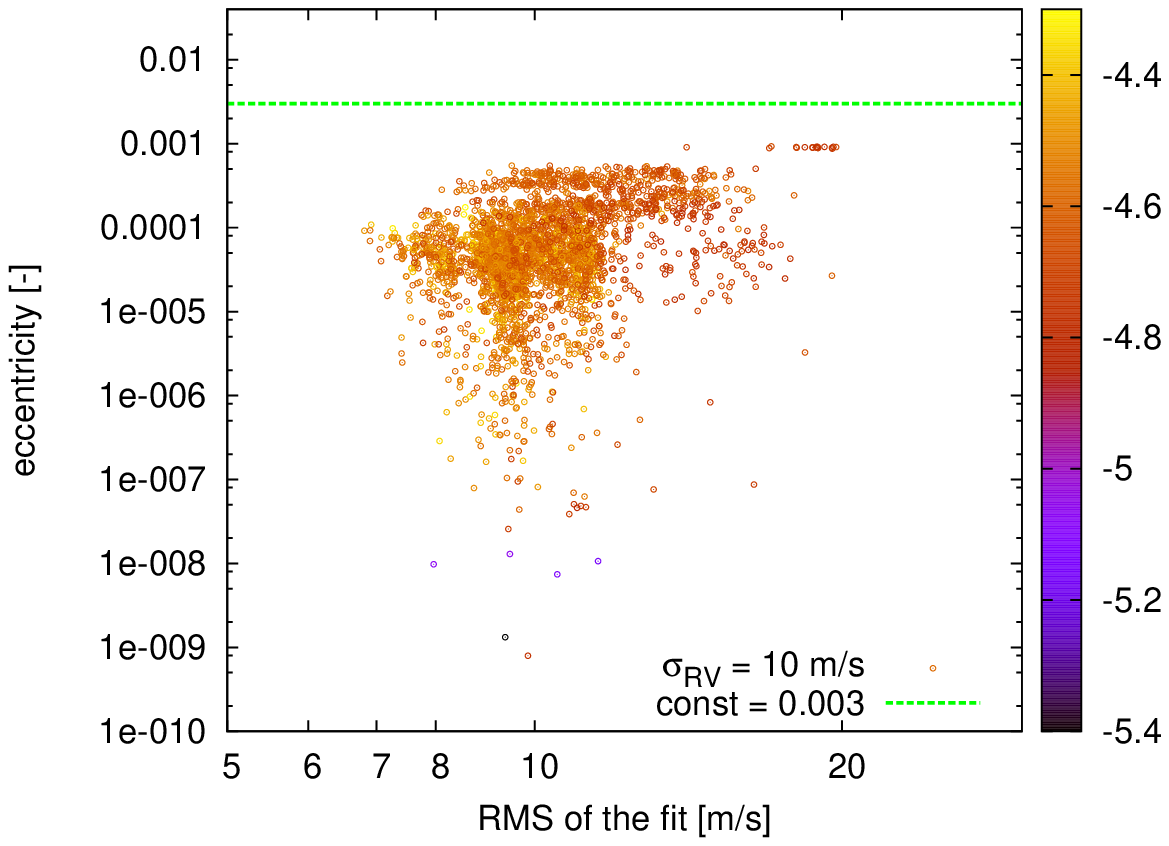}
}
\subfigure[]{
\includegraphics[width=\columnwidth]{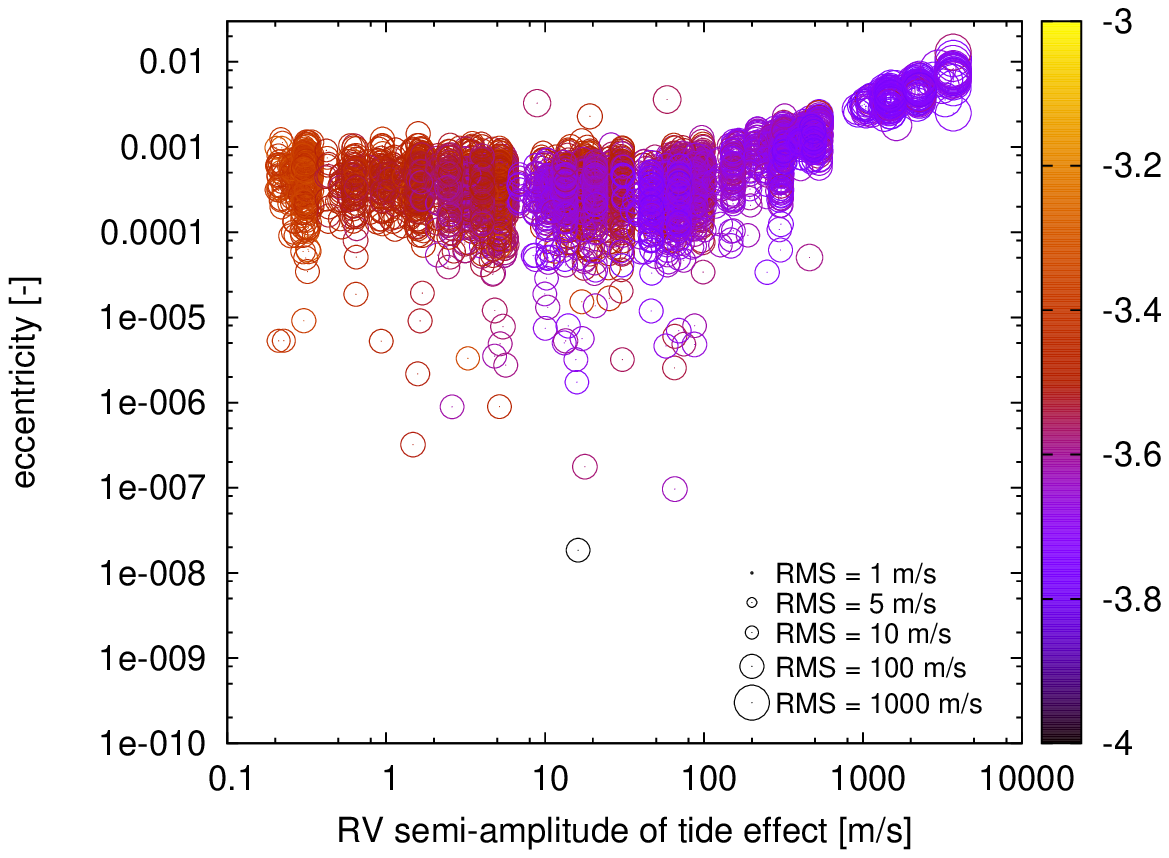}
}
\subfigure[]{
\includegraphics[width=\columnwidth]{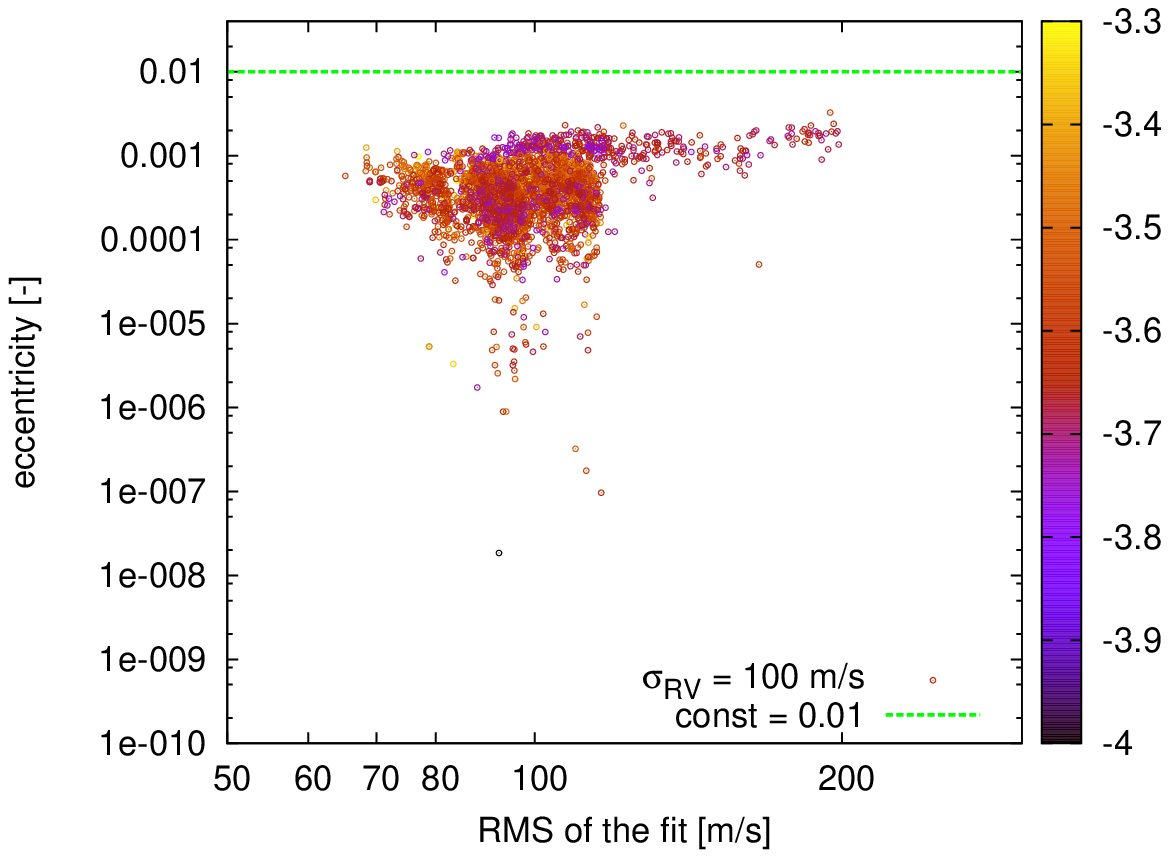}
}
\caption{A more detailed view of the simulations of the tidal effect. The panels (a, c, e) present  all the results from the synthetic set for the three different values of white noise, $\sigma_{RV}=$ 1, 10 and 100 ms$^{-1}$ respectively. Each panel shows the relation between the highest out of two components semi-amplitude of the tidal effect and the best-fit eccentricity. The corresponding figures in the right column are the results of the simulation with $2\sigma_{RV}$ limit applied. The size of the symbol in the first column is defined by the best-fit RMS given by $\log{\textrm{RMS}}$. The size of the symbol in the second column, panels (b, d, f), is constant. Colour denotes the error of the best-fit eccentricity $\sigma_{e}$ as $\log{\sigma_{e}}$. Results clearly show that the eccentricity may be incorporated into the simple model and no hint of an inadequate model is visible. With the increasing semi-amplitude of the tidal effect, increases the best-fit RMS, $\sigma_{fit}$. One may notice that the semi-amplitude even a few times bigger than the white noise $\sigma_{RV}$, returns a reasonable results with low RMS and low $\sigma_{e}$. It may be hard to distinguish errors coming from the underestimated RV errors and a too simplistic model. It is worth noticing that as the best-fit RMS starts to grow the error of best-fit eccentricity decreases.}
\label{fig:etrms}
\end{figure*}

Figure \ref{fig:etrms} shows the results of simulations and three cases where the best-fit RMS, $\sigma_{fit}$, is within a reasonable limit of $2\sigma_{RV}$ but the fit still returns a non-zero eccentricity. These three cases, Figure \ref{fig:etrms} (b, d, f), illustrates the limitation of a simple Keplerian model, applied to the tidally distorted stars. We also used another marker to rule out most of the synthetic binaries with $e>0$: $e/\sigma_e>3$. The error of the fitted eccentricity $\sigma_e$ is very often high enough to still accommodate zero as a value within $3 \sigma_{e}$ limit. However, for $\sigma_{RV}=$ 1, 10 and 100 ms$^{-1}$,  28\%, 32\% and 10\% of the best-fit $e$, respectively do not meet this criterion. This leaves us with incorrect results that may be treated as a detection of an eccentric orbit. 

We propose to use three safe limits for ruling out eccentric false-positives. They are computed as three times the highest false-positive $e$ from our sample. It is $e=$ 0.0004, 0.003 and 0.01 for the three RV measurements precisions 1, 10 and 100 ms$^{-1}$. The limits are illustrated in Figure \ref{fig:etrms} (b, d, f) via horizontal green lines. One may linearly interpolate given limits, but we do not recommend extrapolation as the simulations were conducted for a specific set of binaries. An eccentricity lower than this limit and even complying with all the restrictions, needs special attention and more advanced model than a simple Keplerian one should be used. Further statistical tests (e.g. \cite{lucy05}) may be used to decide if a more sophisticated model is necessary. It is worth noting that with light curves of eclipsing binaries one may directly measure the eccentricity and argument of periastron. This procedure requires times of minima for both components. Such an eccentricity may be compared with the one derived from RVs.

\section{Relativistic effects and orbital inclination}
\label{sec4}

In the basic Keplerian model, the equations for the RV of the primary star  $V_{\textrm{z}1}$ and the secondary star RV $V_{\textrm{z}2}$, in the direction $\bmath{z}$, are:
\begin{eqnarray}
\label{eq02:simpleRVmodel}
V_{\textrm{z}1,2} = K_{1,2} (\cos{(\omega_{1,2}+\nu)}+e\cos{\omega_{1,2}})+V_{\textrm{z}0}
\end{eqnarray}
where radial velocity semi-amplitudes $K_1$ and $K_2$ are equal to:
\begin{eqnarray}
\label{eq03:simpleRVmodel2}
K_1 = \frac{a 2\pi \sin{i}}{P(q+1)\sqrt{1-e^2}} \\
K_2 = q K_1
\end{eqnarray}
$a$ is the semi-major axis of the component's orbit, $q$ mass ratio of the binary, $P$ stands for period and $e$ is the system's eccentricity. In order to fully describe the system we need the argument of periastron $\omega$, the radial velocity $V_{\textrm{z}0}$ of the center of mass and the true anomaly $\nu$.

The tidal effect is not the only subtle effect that has to be considered while modelling precision RVs of binaries. Relativistic effects come into play often at the level well above the 1 ms$^{-1}$. At least three relativistic contributions have to be considered. The light-time (LT) effect within an orbit and coming from the finite speed of light:
\begin{equation}
\label{eq05:LTE}
\Delta V_{LT 1,2} = \frac{K^2_{1,2}}{c} \sin^2{(\nu+\omega_{1,2})}(1+e\cos{\nu})
\end{equation}
This is a correct approximation to the $\beta^2$ order, where $\beta \equiv v/c$.  The transverse Doppler effect:
\begin{equation}
\label{eq06:TD}
\Delta V_{TD 1,2} = \frac{K^2_{1,2}}{c\sin^2{i}} \left(1+e\cos{\nu}-\frac{1-e^2}{2}\right)
\end{equation}
and the gravitational redshift:
\begin{equation}
\label{eq07:GR}
\Delta V_{GR 1,2} = \frac{K_{1,2}(2K_{1,2}+K_{2,1})}{c\sin^2{i}}e
\end{equation}
The speed of light is denoted by $c$. One may find more details on those equations in the papers by \cite{konacki10} or \cite{zucker07}. The principals of the relativistic effects on the motion of a binary star are well described by \cite{kopeikin99}. We omit other relativistic effects of orders higher than $\beta^2$.

\begin{figure*}
\centering
\subfigure[]{
\includegraphics[width=\columnwidth]{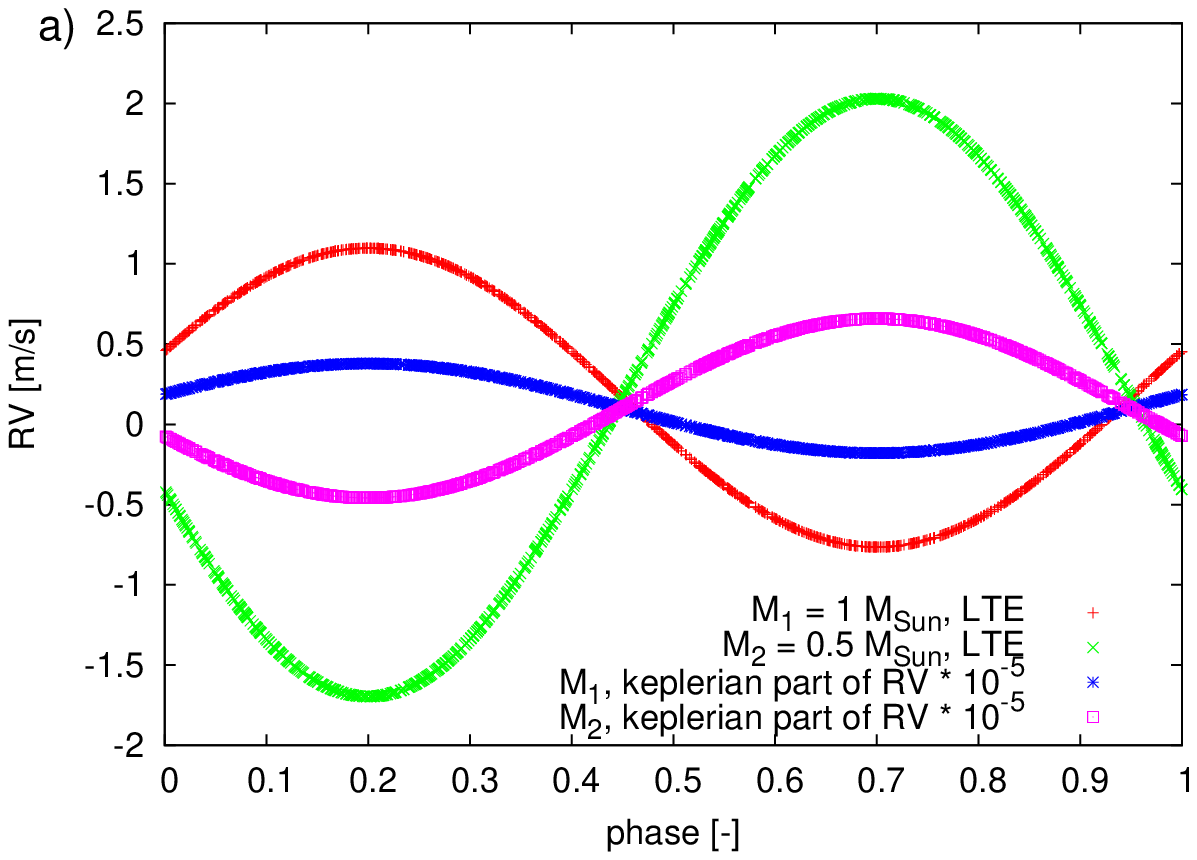}
}
\subfigure[]{
\includegraphics[width=\columnwidth]{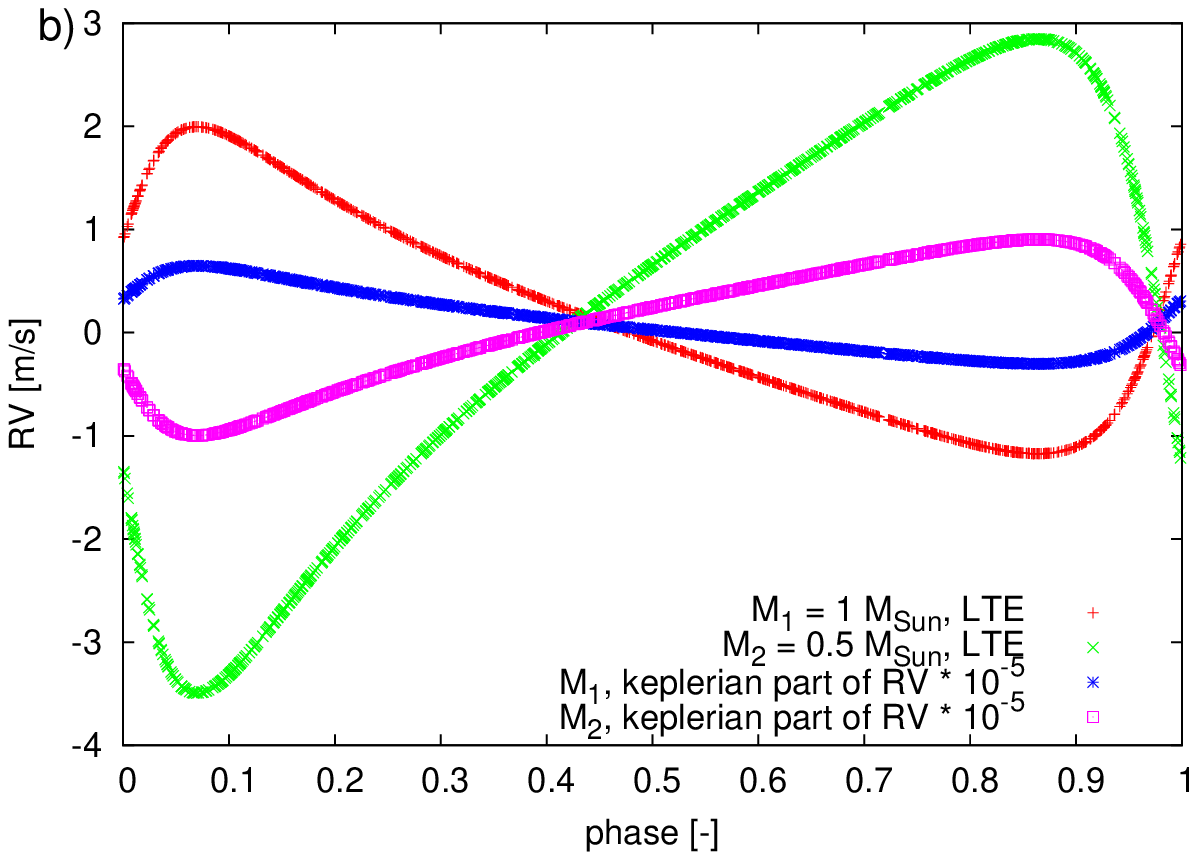}
}
\subfigure[]{
\includegraphics[width=\columnwidth]{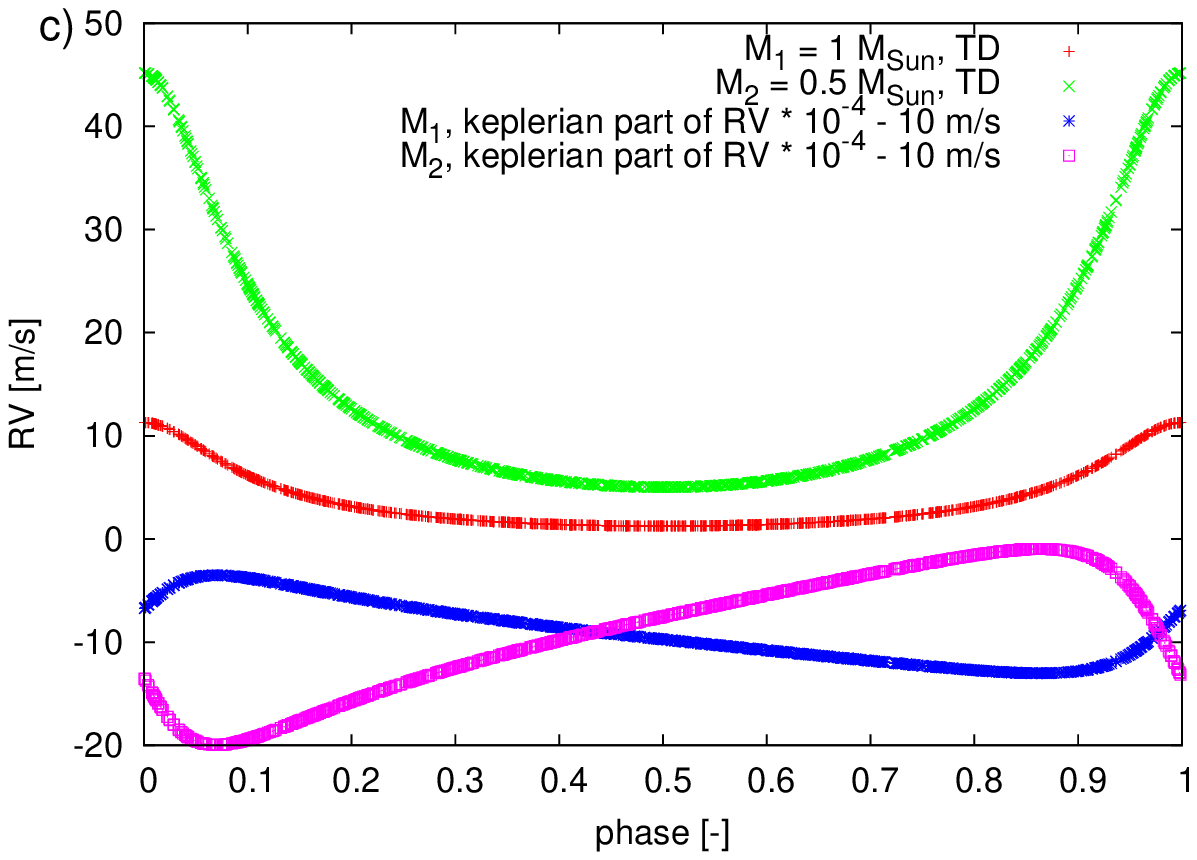}
}
\subfigure[]{
\includegraphics[width=\columnwidth]{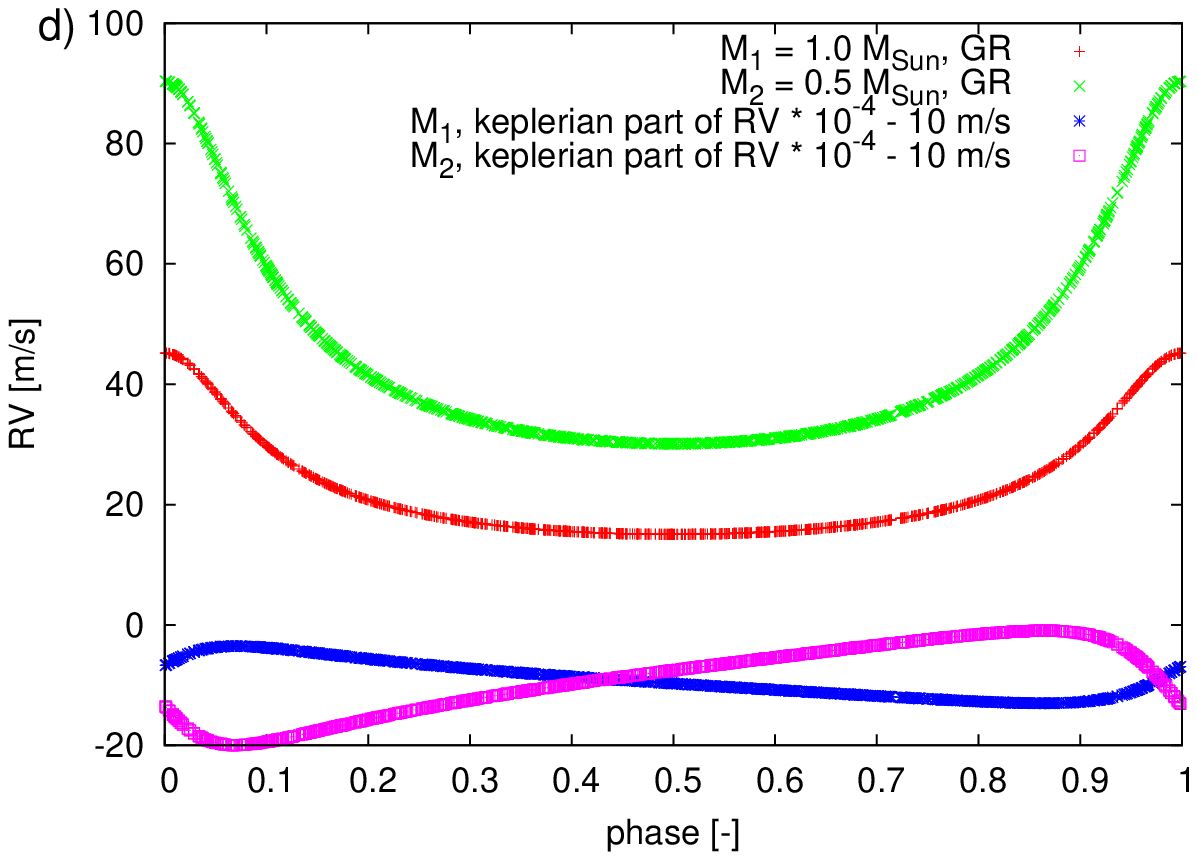}
}
\caption{Two binary systems are shown. The first one is presented in the panel (a) and the other one in the panels (b) - (d). They have different eccentricities of 0 and 0.2. The inclination is 29$\degr$ and 40$\degr$, respectively, for the first and the second system. The common parameters for both binaries are the mass of the first and the second component of 1 and 0.5 M$_\odot$, $\omega=$ 108$\degr$ and period $P=$ 5 days. The most important difference comes from a non-zero $e$ in the second case. Because of this the transverse Doppler effect and the gravitational redshift change with time. The Keplerian part of the RV curve is scaled and shifted for clarity of the figures. The semi-amplitude of the effects is quite high, however it is not unusual. From our sample we estimate that the minimal difference between the Keplerian RV curve and the curve with all three additional effects to be of the order of few meters per second. The average difference is equal to 58.62 ms$^{-1}$ and in the extreme case of high eccentricity binaries, can go as high as a few hundreds meters per second.}
\label{fig:relEff}
\end{figure*}

We have used Equations (\ref{eq05:LTE}) - (\ref{eq07:GR}) to generate a set of RVs of over 400000 synthetic close binaries and explored the average scale and impact of the above effects on the RV curves. In our illustration we chose two systems with different eccentricities. In the first case where only LT effect is visible $e=$ 0 and in the second case $e=$ 0.2. In the first case TD and GR effects are constant but in the second case vary with time. Other parameters as well as images illustrating the contribution of each effect to the final RV are shown in Figure \ref{fig:relEff}. Our main interest is however in a way to obtain the inclination of an orbit purely from RV measurements, via equation \citep{zucker07}:
\begin{equation}
\label{eq06:inclination}
\sin^2{i}=\frac{3e}{\omega'_2-\omega'_1}\frac{K'_{S2}+K'_{S1}}{c}
\end{equation}
where
\begin{equation}
\label{eq06:inclination2}
\omega'_j = -\arctan{\left( \frac{K'_{S_j}}{K_{C_j}} \right)}
\end{equation}
The new elements $K_{S1}$, $K_{S2}$, $K_{C1}$ and $K_{C2}$ are the modified elements $K_{1}$ and $K_{2}$ from the simple RV Equation \ref{eq02:simpleRVmodel}, now accommodating relativistic effects. A detailed description and interpretation of these new elements is in \cite{zucker07}.

\begin{figure*}
\centering
\subfigure[]{
\includegraphics[width=\columnwidth]{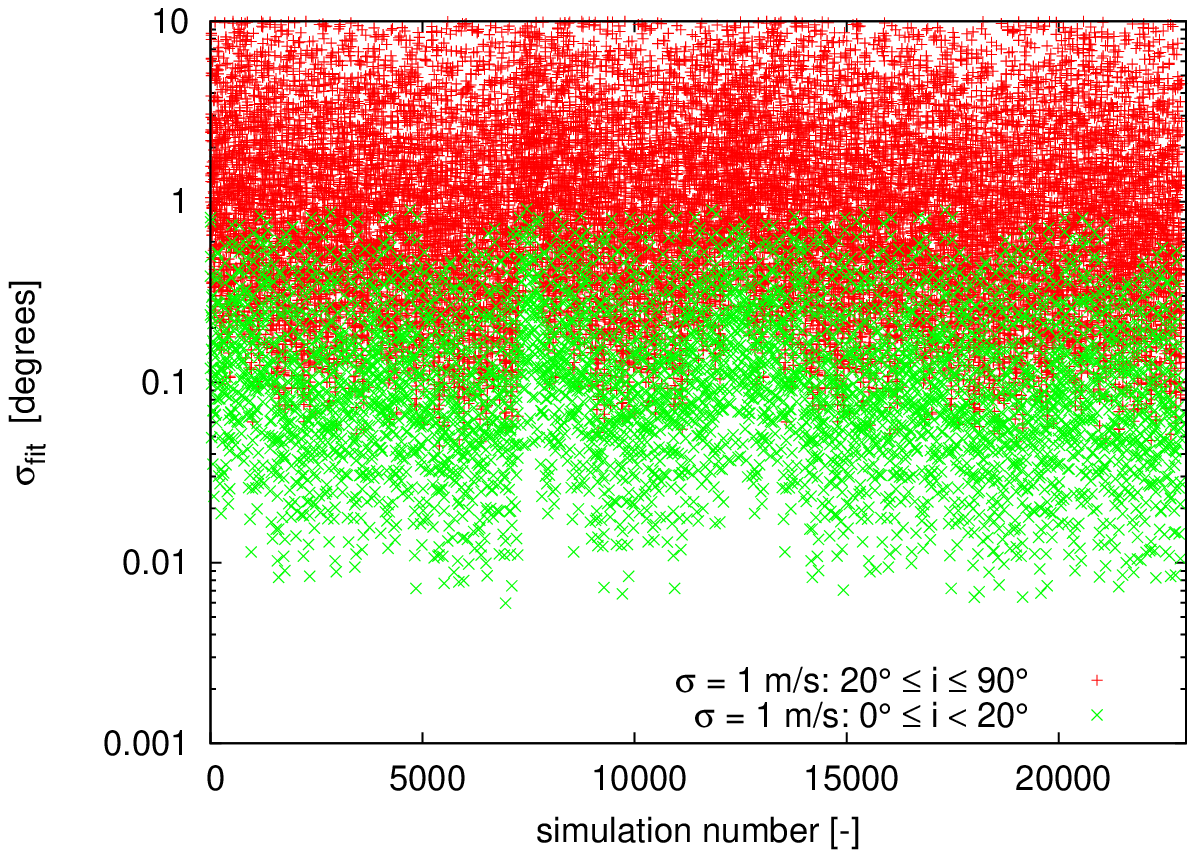}
}
\subfigure[]{
\includegraphics[width=\columnwidth]{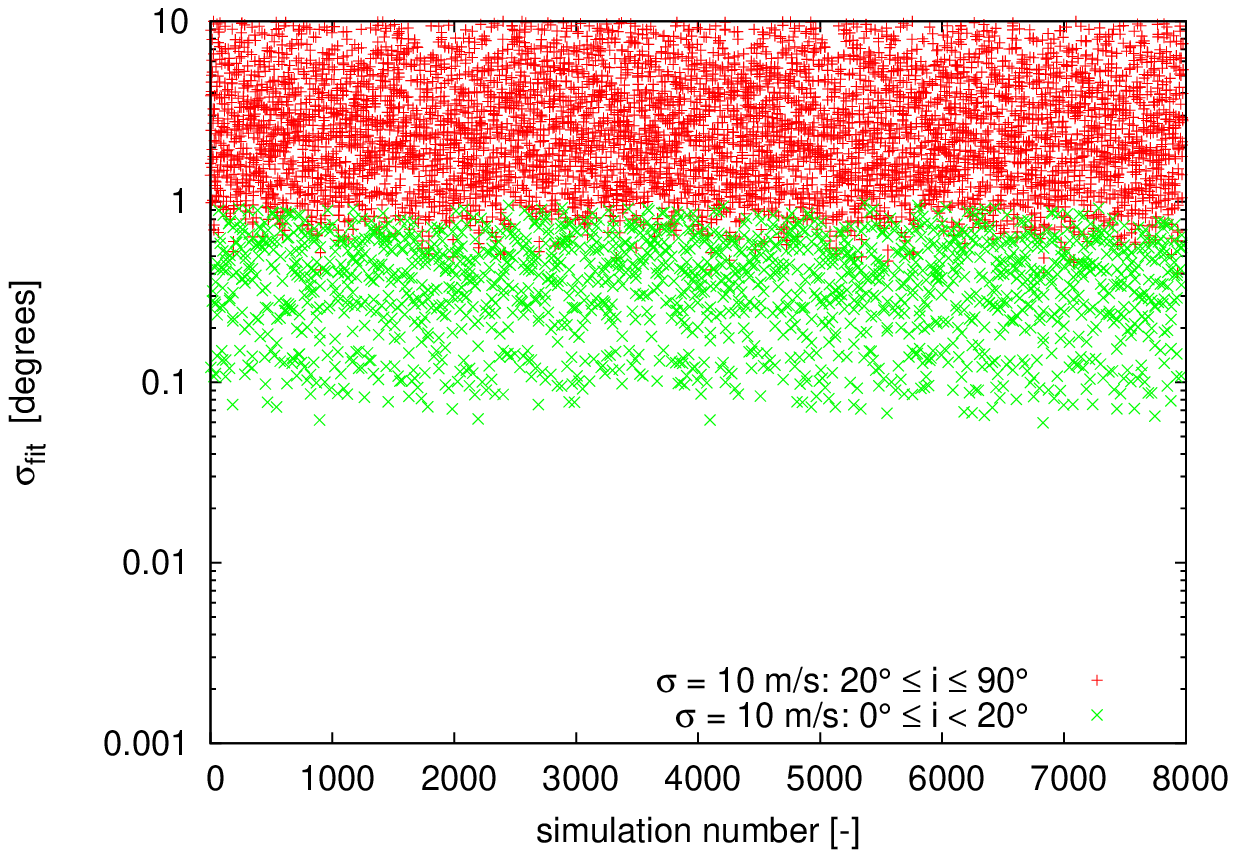}
}
\caption{Two figures showing all the systems where the inclination could be measured, based on thirty simulated RVs for each component of a binary star. The measurement points are randomly chosen from the whole orbital phase. The eclipses are avoided.  All three conditions apply, according to the Set 1 parameters from Table \ref{tab:04_incTresholdsA}. There are no results for $\sigma_{\mathrm{RV}}=$ 100 ms$^{-1}$. Two sets of binaries are distinguished, one (with inclinations lower than 20 degrees) allows much higher precision in inclination measurement, but such systems will be hard to observe with high RV precision due to the spectral lines blending. The second set($i\leq 90$ degrees and $i\geq 20$) degrees is more likely to return valid measurements in real observing conditions.}
\label{fig:04_incP110}
\end{figure*}

In these simulations we assumed the three following conditions:
\begin{itemize}
\item The total RMS of fitted relativistic model is lower than three times the white noise applied to the RV data: $\sigma_{fit}<A \sigma_{\mathrm{RV}}$, where $A=$3.
\item The error of inclination  $\sigma_i$ is ten times lower than inclination $i$: $i > B \sigma_i$, where $B=$ 10.
\item The difference between the true inclination $i_t$ and the best-fit inclination is lower than $3\sigma_i$: $|i_t-i|<C\sigma_i$, where $C=$3.
\end{itemize}
The result of such a simulation is presented in Figure \ref{fig:04_incP110}. It clearly shows that the proposed method works, but obviously only for eccentric orbits. The precision of the fit can be as good as 0.01$\degr$ for 1 ms$^{-1}$ accuracy of the RV data and 0.1$\degr$ for 10 ms$^{-1}$ accuracy. For 100 ms$^{-1}$ precision there is no single system in the final pool. However these values are possible only for very low inclinations, lower than 20 degrees. Such systems will have spectral lines blended, making the measurement of RV a very hard or even impossible task. Altogether it means that 40\% and 25\% of results of respective 1 ms$^{-1}$ and 10 ms$^{-1}$ precisions meet the conditions. We are in a good situation because we know the true inclination. However removing the third condition does not change the whole picture and adds only a fraction of percent to the results. When we loosen our three conditions we get a higher percentage of acceptable results. \textbf{Table \ref{tab:04_incTresholdsA} includes eight different examples. First four use the set of criteria described above and for the rest we used the modified second criterion: $|\omega'_1-\omega'_2| > B  \sigma_{\omega'_1-\omega'_2}$. The results for that set are denoted by an apostrophe. The case described above is Set 1. It is worth noticing that both criteria give very similar results but from the mathematical point of view $\omega'_1$ and $\omega'_2$ may be better indicators of ability to measure relativistic effects.} 

\begin{table}
\caption{The impact of different thresholds on the number of acceptable solutions for the orbital inclination. The columns 5-7 represent a percent of systems from the synthetic set where all three conditions with respective $A$, $B$ and $C$ values were met. The typical number of systems per one $\sigma_{\mathrm{RV}}$ value and per one set is 51000, where 10\% of them have a circular orbit. The relativistic effect is much more sensitive to small inclinations than to large (edge-on orbits) inclinations. The effect strongly depends on the ability to detect differences between $\omega'_2$ and $\omega'_1$. }
\label{tab:04_incTresholdsA}
\centering
\begin{tabular}{r c c c c c c}
\hline
Parameter Set & $A$ & $B$ & $C$  & 1 ms$^{-1}$ & 10 ms$^{-1}$ & 100 ms$^{-1}$  \\
\hline
\hline
Set 1 & 3 & 10 & 3 & 40.1\% & 25.0\% & 0.0\% \\
Set 2 & 3 & 5 & 3 & 45.3\% & 43.0\% & 0.7\% \\
Set 3 & 3 & 3 & 3 & 48.3\% & 54.7\% & 7.3\% \\
Set 4 & 4 & 2 & 5 & 50.0\% & 64.0\% & 20.5\% \\
Set 1' & 3 & 10 & 3 & 39.5\% & 14.8\% & 0.0\% \\
Set 2' & 3 & 5 & 3 & 43.5\% & 35.7\% & 0.0\% \\
Set 3' & 3 & 3 & 3 & 47.1\% & 47.8\% & 0.5\% \\
Set 4' & 4 & 2 & 5 & 49.8\% & 57.8\% & 5.5\% \\
\hline
\end{tabular}
\end{table}

In order to increase our chances of determining the inclination just from an RV data set, we need to focus on eccentric binaries. With circular orbits it is not possible at all. Higher eccentricity means higher contribution of relativistic effects. With a given binary we may try to increase the number of RV measurements above 30, constant in our set, and to choose moments when spectra are taken. With the numerical model we may easily determine where the deviation from the Keplerian orbit is the highest, and by this increase the efficiency of observations.

\section{Apsidal precession and RV accuracy}
\label{sec5}
\begin{figure*}
\centering
\subfigure[]{
\includegraphics[width=\columnwidth]{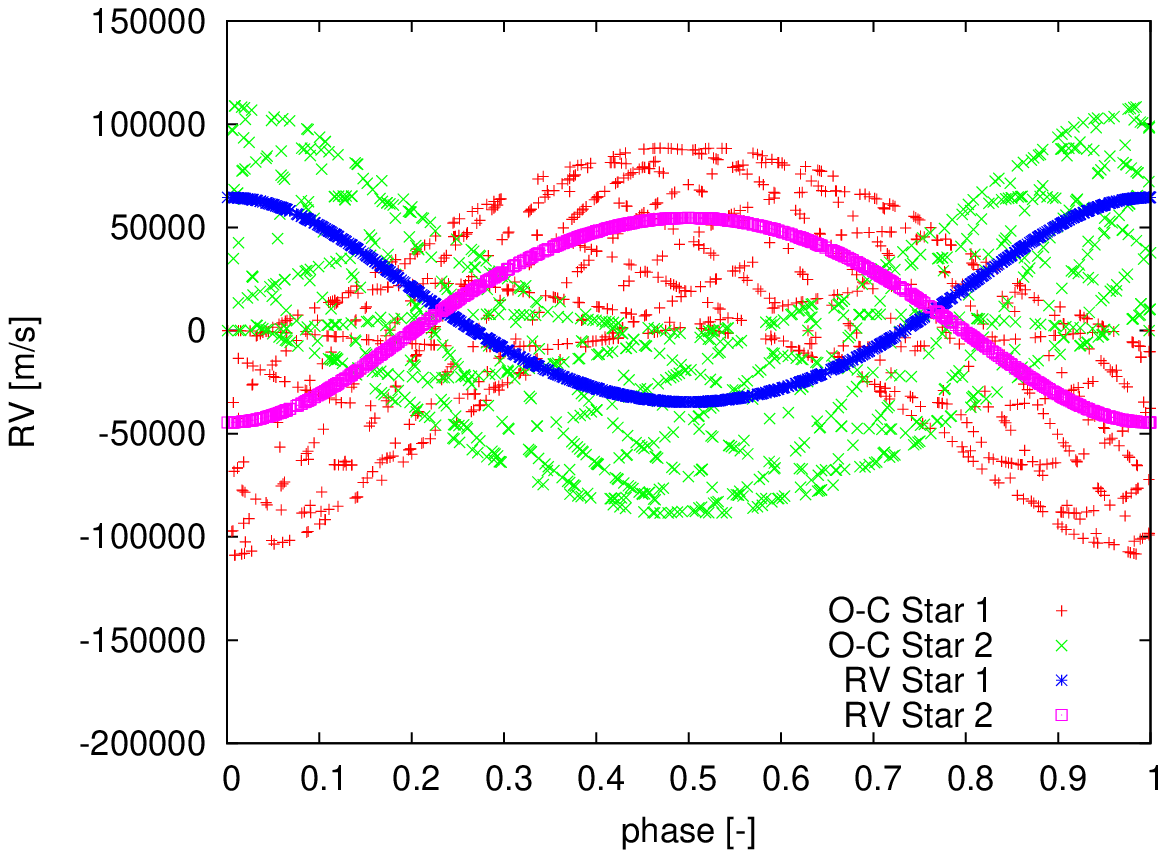}
}
\subfigure[]{
\includegraphics[width=\columnwidth]{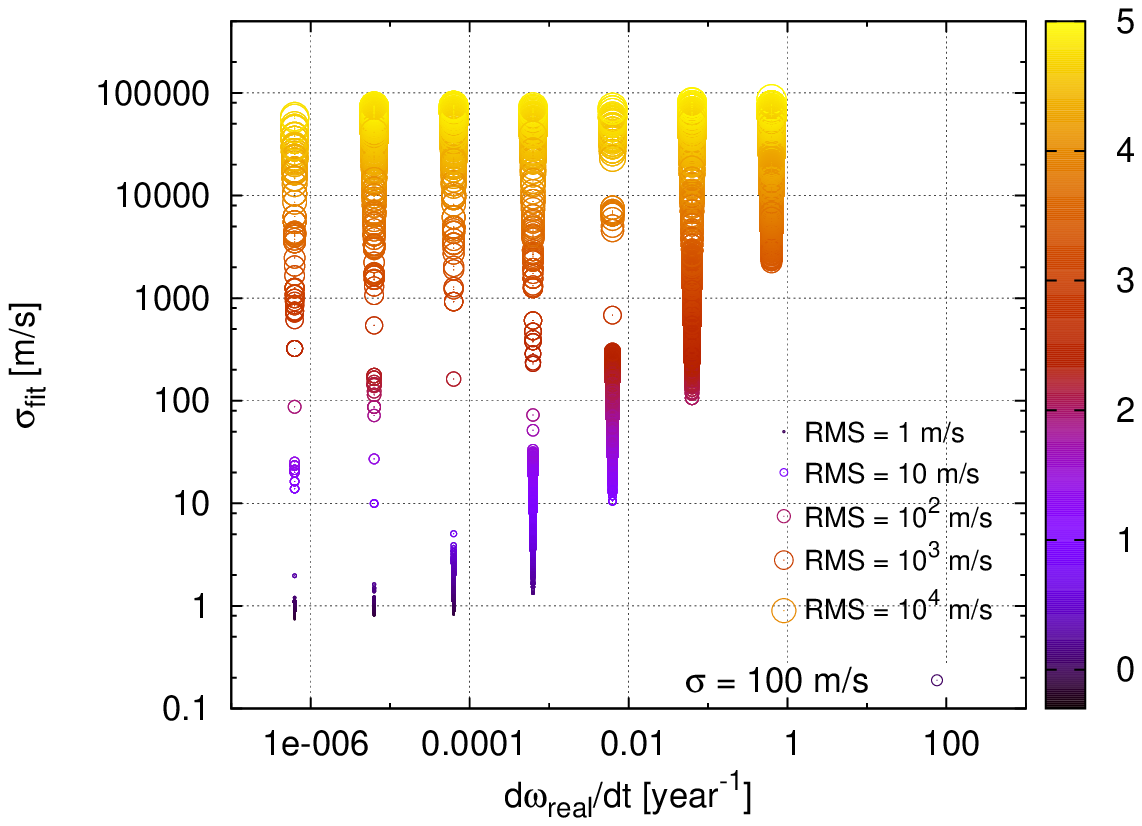}
}
\subfigure[]{
\includegraphics[width=\columnwidth]{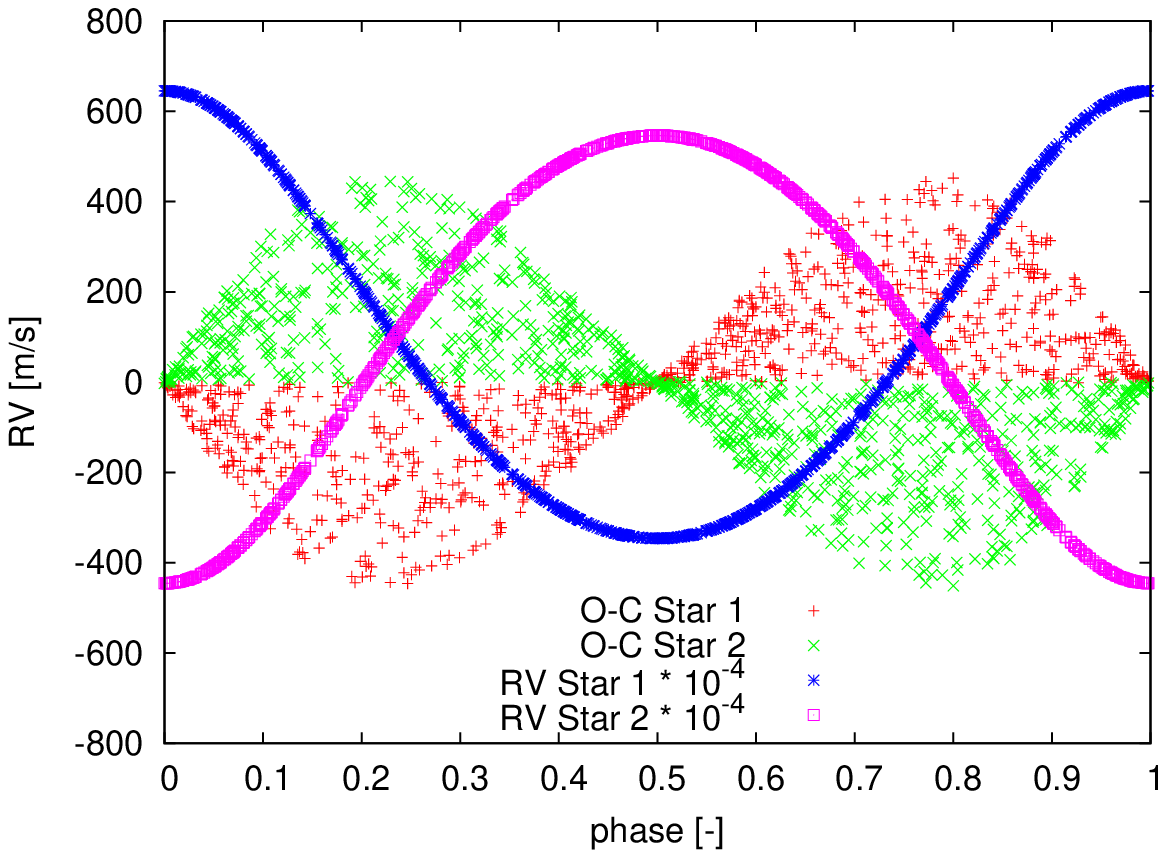}
}
\subfigure[]{
\includegraphics[width=\columnwidth]{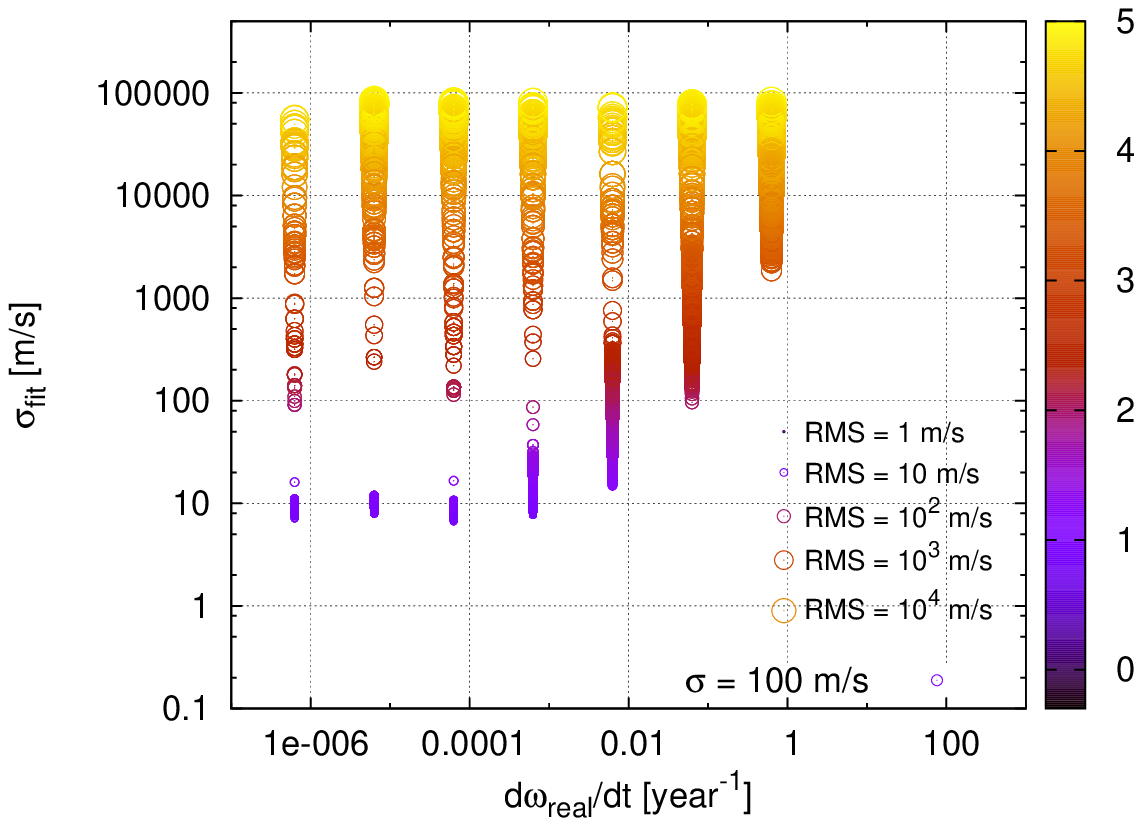}
}
\subfigure[]{
\includegraphics[width=\columnwidth]{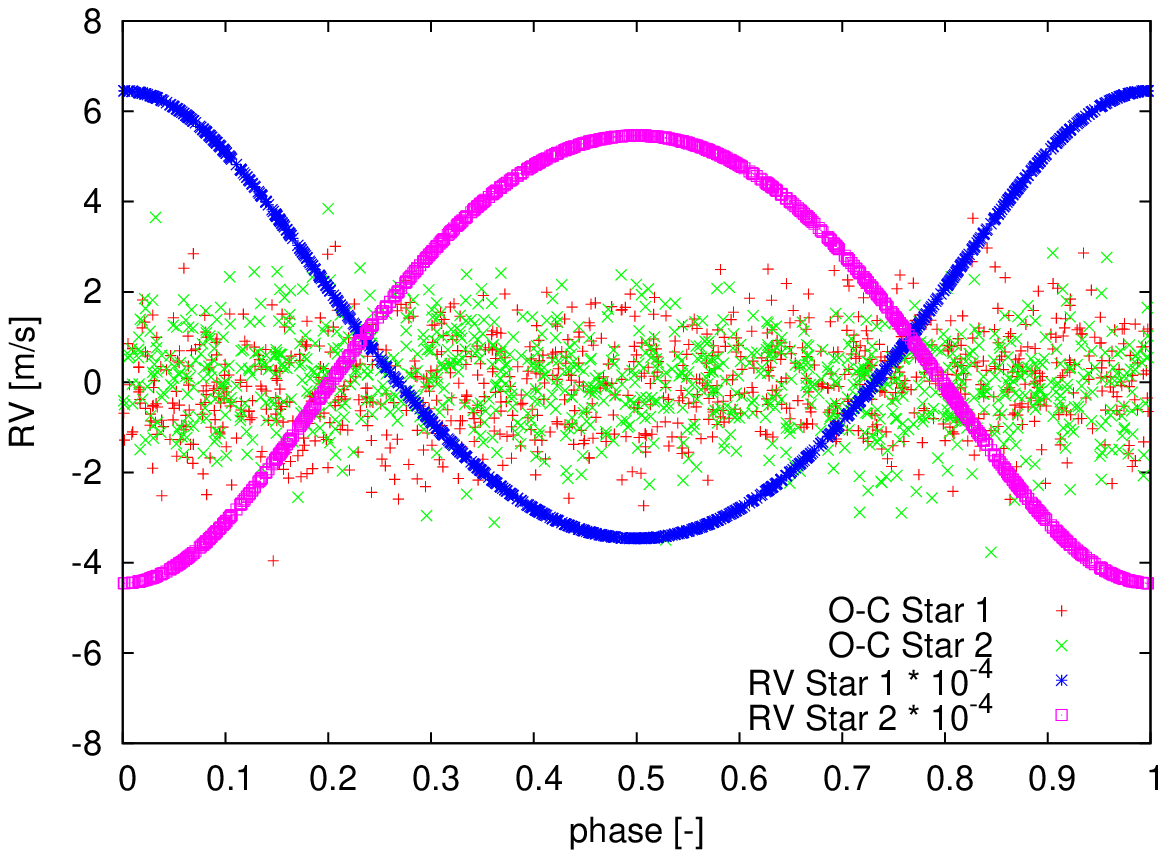}
}
\subfigure[]{
\includegraphics[width=\columnwidth]{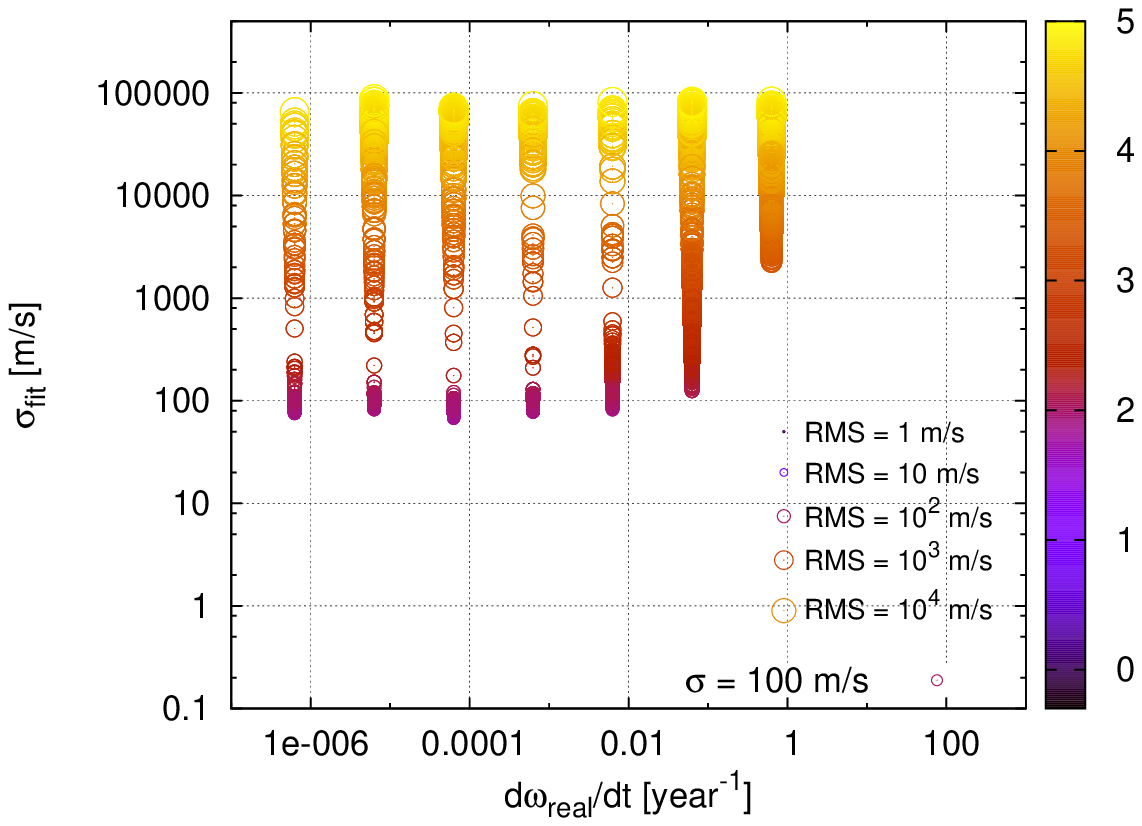}
}
\caption{The left column, panels (a), (c) and (e), contains examples of simple RV curves of both components with $d\omega/dt=$ 0. When we add $d\omega/dt=$ $10^{-1}$, $10^{-5}$, $10^{-8}$ to those three different RV curves, we can calculate the difference between data with and without an apsidal motion. In the first case the difference is huge, and from the shape of O-C we may deduce an apsidal motion in the system. In the panel (c) it is still visible, but in the case of (e) it may be easily missed and incorporated into the total white noise. In the right column, panels (b), (d) and (f), two major groups of results may be distinguished: one group with a high $\sigma_{fit}$ comes from the fact that a random distribution of input parameters and measurements may result in a failure of the least square method. However, most of the systems are in the second group around the value of the white noise $\sigma_{\mathrm{RV}}$ for low $\omega_{real}/dt$. At some point, as the $\omega_{real}/dt$ increases, the results start to drift off the constant line in the direction of a higher RMS of the fit. The colour and the size of circles denotes $\log{\sigma_{fit}}$.}
\label{fig:domegaEff}
\end{figure*}

The apsidal precession is an important parameter, having great impact on the RV curve. The main cause of this effect may be tidal or relativistic precession. A misalignment between the star's spin and the orbital angular momentum of an eclipsing binary star or a third body may cause precession as well \citep{mazeh08}. To precisely measure the orbital and physical parameters of binaries we have to take into account this effect. In the case of eclipsing binaries, it may also be used to confront the models of stellar structure with observations via the stellar radial concentration parameter \citep{mazeh08}. Three examples of $d\omega/dt=$ $10^{-1}$, $10^{-5}$, $10^{-8}$ rad yr$^{-1}$ are presented in the left-hand column of Figure \ref{fig:domegaEff}. In our simulations we used units of radians per year to describe the effect. This approach is insensitive to the different periods of systems used in the simulations. The pool of synthetic binaries consists of 70529 objects, and thirty simulated RV measurements for each component are randomly distributed over the four year time span. Only one inclination, 93$\degr$, was chosen. The ranges of the starting values of $\omega$ and $e$ are presented in Table \ref{tab:03_WdParameters}.

\begin{figure*}
\centering
\subfigure[]{
\includegraphics[width=\columnwidth]{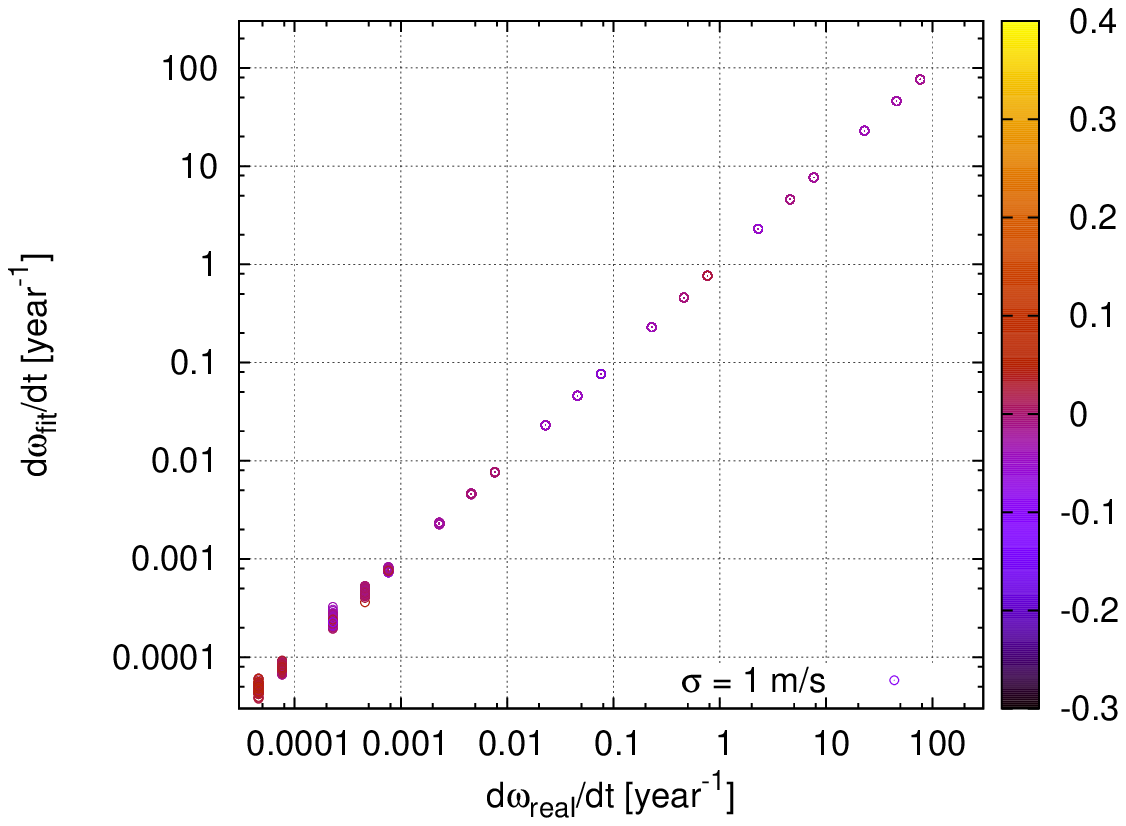}
}
\subfigure[]{
\includegraphics[width=\columnwidth]{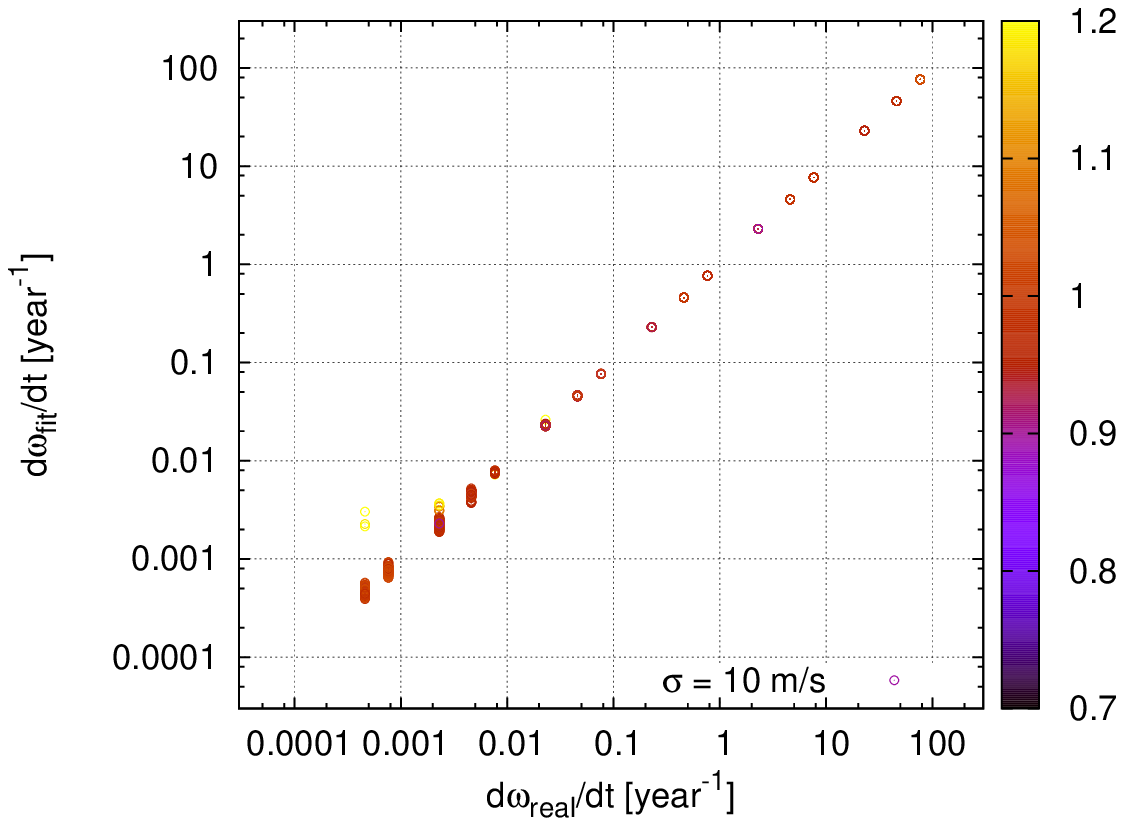}
}
\caption{A comparison of $\omega_{real}/dt$ and $\omega_{fit}/dt$ for 1 ms$^{-1}$  (left) and 10 ms$^{-1}$ precision (right). The figure for 100 ms$^{-1}$ precision is not presented here. It is very similar to the two results shown here, but with correct results starting around $\omega_{real}/dt=$ 0.01. It is consistent with the conclusions from Figure \ref{fig:domegaEff} (f). Two conditions similar to these in Section \ref{sec4} are applied to the results. The parameters $A=$ 2 and $B=$ 10 are conservative but guarantee correct results. The colour denotes $\log{\sigma_{fit}}$, where $\sigma_{fit}$ is given in radians.}
\label{fig:05_dog110}
\end{figure*}

The results of our simulations are presented in Figures \ref{fig:domegaEff} and \ref{fig:05_dog110}. In the first figure the data used in simulations contained apsidal precession, but only the Keplerian model was fitted. In the second case we fitted an additional parameter $d\omega_{fit}/dt$ and compared the results of the best fit values with the real value of the parameter $d\omega_{real}/dt$. In this simulation we assumed two conditions similar to the ones used in Section \ref{sec4}. The parameters used were $A=$ 2 and $B=$ 10.

It is important to note the threshold where the apsidal motion can no longer be absorbed by a simple Keplerian model and requires an additional parameter to describe observations. It is $\omega_{real}/dt=$ $10^{-4}$, $10^{-3}$, and $10^{-2}$ for $\sigma_{\mathrm{RV}}=$ 1, 10 and 100 ms$^{-1}$ respectively. The threshold was estimated with simulations and is based on the systems meeting the following criteria:
\begin{itemize}
\item The total RMS of fitted Keplerian model is lower than three times the white noise applied to the RV data.
\item The error of $d\omega$ is ten times lower than $d\omega$.
\end{itemize}

A recent paper by \cite{konacki10} includes five binaries where the RV precision reached the level of 2-10 ms$^{-1}$. Four objects showed no apsidal motion, so with our simulations we may say that $d\omega/dt=$ $10^{-3}$ rad yr$^{-1}$ is ruled out in those systems. However one system, HD78418, showed a small value of the apsidal motion. It is (1.5 $\pm$ 0.4)$10^{-3}$ rad yr$^{-1}$, still on the edge of detectability. Those results comply with the conclusions of our simulations connecting the precision of RV measurements with the limits on $d\omega/dt$.

\begin{figure}
\includegraphics[scale=0.65]{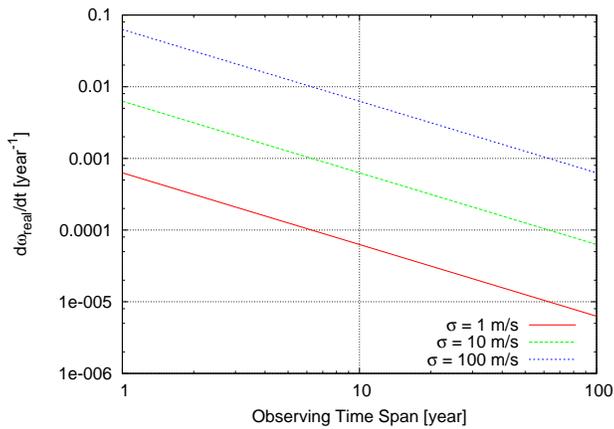}
\caption{The relation between the observing time span and detectable $\omega_{real}/dt$ for the three different RV measurement precisions. The true limit for a particular system may be even a bit lower, as this is an upper estimation of the limit. However we expect all the systems with $\omega_{real}/dt$ observed with sufficient RV precision to yield the correct, non-zero $d\omega/dt$.}
\label{fig:omegaGraph}
\end{figure}

We checked the relation between the simulated time span of the observations and ability to detect apsidal motion. With six different time spans: 1, 2, 3, 4, 10 and 100 years, we established a simple relation presented in Figure \ref{fig:omegaGraph}. It is an upper limit for the systems that will show the apsidal motion within the time span of the survey. The limits are based on the group of simulated objects fully meeting the two criteria described earlier. We decided to use bigger groups of results rather than a few single objects as the simulations run on a non-continuous set of input parameters.

\section{Conclusions}
\label{sec6}
Modern precision of RV measurements of binary stars reached a point where the subtle effects coming from the tidal distortion, relativistic effects and apsidal precession can no longer be ignored. The majority of synthetic close binaries we examined required corrections for those effects. A recent paper by \cite{konacki10} shows an RV precision of up to 2 ms$^{-1}$. With this level of precision, we must model the impact of tidal distortions of components of binary stars on the RV curves. When this effect is ignored it may mimic small eccentricities in otherwise circular orbits. On the other hand, the relativistic effects not only need to be accounted for at such an RV precision, but also can be used to determine the inclinations of non-eclipsing binaries based on the RV data only. In favourable cases, the precision of $i\pm$ 0.1$\degr$ should be possible. Finally, we demonstrate that with precision
RVs we can easily detect an apsidal motion lower than $d\omega/dt=$ $10^{-4}$ rad yr$^{-1}$, and in this way contribute to the understanding of dynamics and internal structure of the components of binary stars. 

\section*{Acknowledgments}
We are grateful for all the comments and suggestion made by the reviewer, they improved the overall quality of the paper. This work is supported by the Polish National Science Center grants 2011/03/N/ST9/03192, 5813/B/H03/2011/40 and 2011/03/N/ST9/01819, and by the European Social Fund within an individual project of the Kuyavian-Pomeranian Voivodship scholarship for Ph.D. students ``Krok w przysz{\l}o\'s\'c - III edycja". K.G.H. acknowledges support provided by the Proyecto FONDECYT Postdoctoral No. 3120153. M.K. acknowledges support from the European Research Council via a Starting Grant and the Foundation for Polish Science via a grant "Idee dla Polski".

\bibliographystyle{mn2e}

\bsp

\label{lastpage}

\end{document}